\documentclass{ar-1col}
\usepackage[numbers]{natbib}
\usepackage{url}
\usepackage[export]{adjustbox}
\usepackage{subcaption}
\jname{Annu. Rev. Nucl. Part. Sci.}
\jvol{72}
\jyear{2022}
\doi{10.1146/annurev-nucl-111119-041046}

\begin{document}

\markboth{M.S. Turner}{Precision Cosmology}

\title{The Road to Precision Cosmology}

\author{Michael S. Turner
\affil{Kavli Institute for Cosmological Physics, University of Chicago, Chicago, IL, USA, 60637-1433; email: mturner@uchicago.edu}
\affil{The Kavli Foundation, Los Angeles, CA, USA, 90230-6316}}

\begin{abstract}

The past 50 years has seen cosmology go  from a field known for the errors being in the exponents to precision science.  The transformation,  powered by ideas, technology, a paradigm shift and culture change, has revolutionized our understanding of the Universe, with the $\Lambda$CDM paradigm as its crowning achievement. 
I chronicle the journey of precision cosmology and finish with my thoughts about what lies ahead.
\end{abstract}


\maketitle

\tableofcontents

\section{OVERVIEW}
Only a half century ago cosmology was the province of west-coast astronomers, the total number of redshifts measured was a few hundred and high redshift meant $z\sim 0.1$.  
A ``standard model" -- the hot big bang model -- had just emerged.   Named and described in detail  by physicist Steven Weinberg, it took the Universe from 
an early hot soup of hadrons to 
galaxies expanding away from one another today.

Today, cosmology is mainstream, ``industrial" science with a mix of more than a thousand astronomers and physicists around the world working to both understand the birth and evolution of the Universe and to learn about the fundamental nature of matter, energy, space and time.  It used to be said about cosmology that the errors are in the exponents; true or not, the field was certainly data starved.  
Today, the term precision cosmology 
is a reality, and cosmology is an exemplar of big-data science.

The current paradigm, $\Lambda$CDM, describes the Universe from a fraction of a second after the big bang, when seeds for all the structure we see today were quantum fluctuations on subatomic scales, to the present, $13.80\pm 0.023\,$Gyr later.  $\Lambda$CDM is supported by a wealth of cosmological data, from tens of millions of redshifts extending to $z = 10$ to nanoKelvin measurements of the cosmic microwave background on angular scales down to arcminutes.  

The progress over the past 50 years has been remarkable, driven by ideas, technology, big discoveries, a paradigm shift, and culture change. My review 
is not comprehensive (that would require a book), and both benefits and suffers from being the perspective of an active participant.  
While it has all the usual features, I hope that the reader also ``hears" the story of the grand adventure that the past 50 years has been.  

The first section is devoted to beginnings:  the development of the hot big-bang model, the explosion of technology and the entrance of particle physicists into cosmology.  The next  chronicles the development of the $\Lambda$CDM paradigm:  the dark matter paradigm shift, the CMB discoveries that mark the birth of precision cosmology, the two big ideas -- inflation  and CDM -- that 
underpin it, and last, but not least, the discovery of cosmic acceleration and dark energy which completed $\Lambda$CDM. The final section sings the praises of $\Lambda$CDM, discusses some of the challenges ahead, and ends with some personal reflections.




\section{BEGINNINGS}
\subsection{The first paradigm:  the hot big-bang model}
The Universe is big in both space and time and for much of human history it has been largely beyond the reach of our boldest ideas and most powerful instruments.  I mark the birth of modern cosmology at roughly 100 years ago.  Einstein had introduced General Relativity, the first theory of gravity and spacetime capable of describing the entire Universe, and the first cosmological solutions had been found (e.g., the deSitter, Friedmann and Lema{\^ i}tre solutions as well as Einstein's static model).  At about the same time,   Hale and  Ritchey invented the (modern) reflecting telescope, and Hale moved astronomy to the mountain tops of California, first Mt. Wilson and later, Mt. Palomar. With bold ideas and new instruments, astronomers were ready to explore the Universe beyond our own Milky Way galaxy and begin to discover and understand the larger picture.

Hale's second big reflector, the 100-inch Hooker telescope, enabled Hubble to discover that galaxies are the building blocks of the  Universe today and that  it is expanding, the signature of its big bang beginning.  While it took a few years to connect the solutions of General Relativity to the observational data, the basics of the big bang model were in place.

The ``hot" in the hot big-bang model came with  Penzias and  Wilson's discovery of the cosmic microwave background (CMB) in 1964 \cite{PW}.  While the idea of a hot beginning was introduced by Gamow and his collaborators in 1948 \cite{abc}, to explain the non-equilibrium origin of the chemical elements,
the 1964 discovery was accidental, an interesting and oft-told story.\footnote{Both the CMB story and a more detailed history of the discovery and interpretation of the expansion can be found in \cite{PJEPCC}.}

In 1972, years before Standard Model referred to the remarkable theory that describes quarks and leptons, Steven Weinberg coined the term ``standard model" for the hot big bang model and described it in his classic textbook \cite{SWeinberg}.  This standard model traces the Universe from a hot soup of hadrons at around $10^{-5}\,$sec through the synthesis of the light elements (largely $^4$He with traces amounts of D, $^3$He and $^7$Li) at a few seconds to the formation of neutral atoms and the last-scattering of CMB photons at around 400,000 years after the big bang, and finally to the formation of stars and galaxies.

The first paradigm embodied the basic cosmological picture -- expansion from a hot big bang beginning to a Universe filled with galaxies moving away from one another today.  General Relativity, nuclear physics (for big bang nucleosynthesis, or BBN), and atomic physics (for the interpretation of the CMB) provided a strong theoretical foundation.  The triad of the expansion, the light-element abundances, and the blackbody spectrum of the CMB provided an equally strong observational foundation.

\begin{figure}[h]
\begin{subfigure}{0.55\textwidth}
\includegraphics[width=0.95\linewidth, height = 4.5cm]{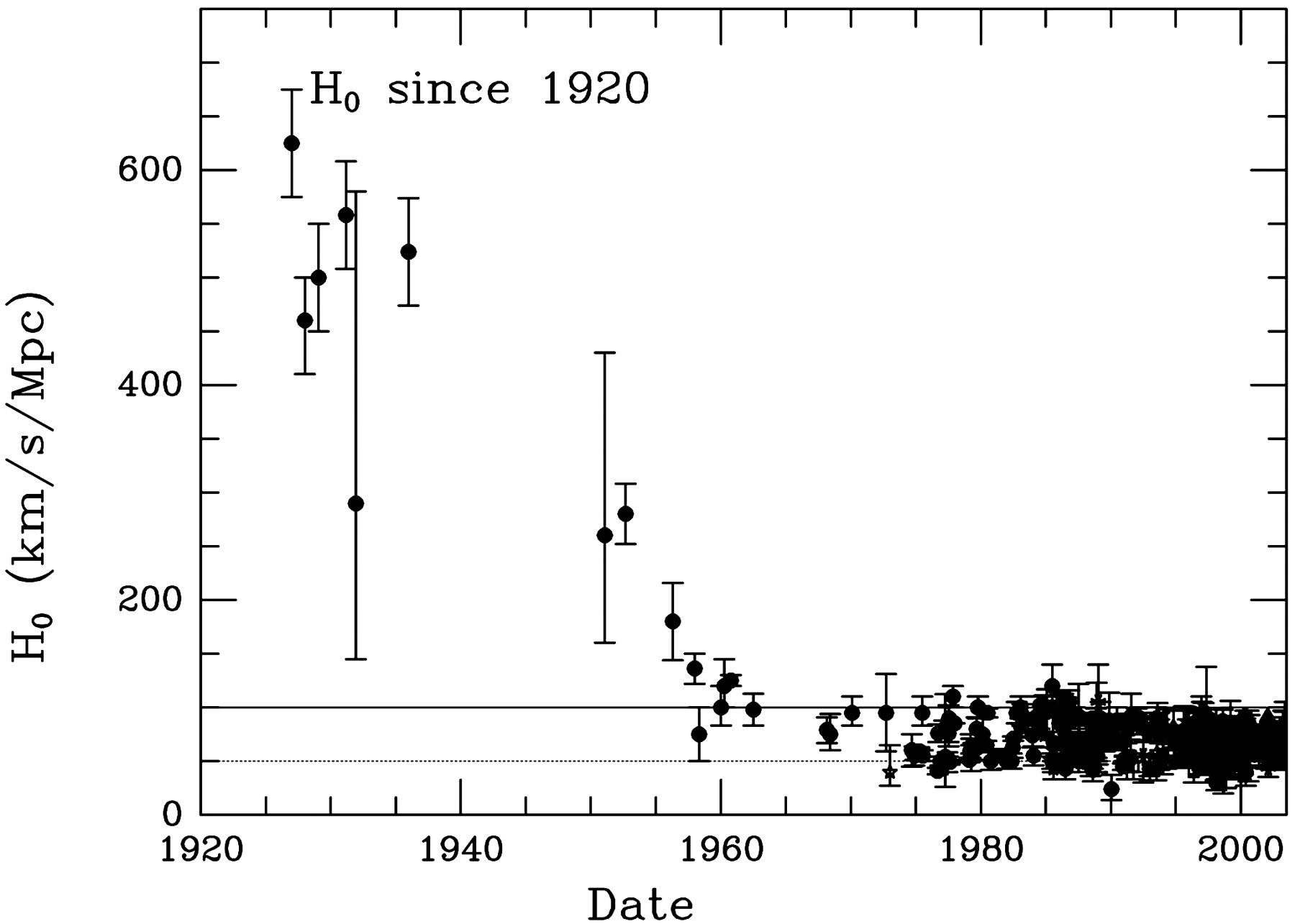}
\end{subfigure}
\begin{subfigure}{0.45\textwidth}
\includegraphics[width=0.95\linewidth, height = 2.75cm]{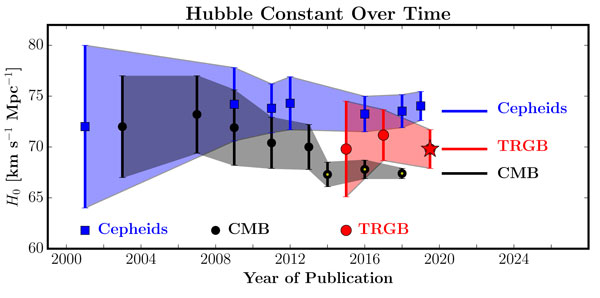}
\end{subfigure}
\caption{``Low-" and ``high-precision" cosmology.  left: $H_0$ measurements from 1920 to 2000.  
By 1970, most measurements were between 50 and 100 km/s/Mpc, but with unrealistically small error bars.  The HST Key Project changed that with its 2000 determination, $H_0 = 72 \pm 2 \pm 6\,$km/s/Mpc.  right: $H_0$ measurements since 2000.   ``Cepheid" and ``TRGB" are direct measurements of $H_0$ and ``CMB" use  CMB measurements and the assumption of the $\Lambda$CDM to determine $H_0$.  Note, the current ``Hubble tension."}
\label{fig:H0}
\end{figure}

In 1970, Sandage summed up cosmology  as the search for two numbers, $H_0$ and $q_0$.  The expansion rate of the Universe $H_0$ also sets the age of the Universe, $t_0 = aH_0^{-1}$, with the deceleration deceleration parameter $q_0$ determining the constant $a$.  And, for a universe comprised only of matter, $q_0$, the ratio of the matter density to the critical density $\Omega_0$, and the curvature radius of the universe are related:  $q_0 = \Omega_0/2$ and $R_c = H_0^{-1}/|\Omega_0 -1 |^{1/2}$.

 It would take until 2000 and the Hubble Space Telescope Key Project to pin down $H_0$ with a reliable error bar:  $H_0 = 72 \pm 2 \pm 6\,$km/s/Mpc (statistical and systematic) \cite{HSTKeyProject}.  As I discuss later, $H_0$ is still a lively topic today and continues to live up to its reputation as the most important number in cosmology (see Fig.~\ref{fig:H0}).

As for the deceleration parameter, the Universe is actually accelerating and $q_0$ cannot actually be measured with both precision and accuracy.\footnote{The reason is simple, $q_0$ is the second term in a Taylor expansion for the distance to redshift $z$; at large enough redshifts to determine $q_0$, the Taylor expansion is not accurate.  There are many better parameterizations for cosmic distance; see \cite{NebenTurner}.}  While its value can be inferred, $q_0 = -0.55 \pm 0.05$, $q_0$ has been replaced by other cosmological parameters that better capture the physics and that can be be measured with accuracy and precision.

Two big questions loomed:  How did structure form? and What happened during the first microsecond when the soup of hot hadrons would have been strongly-interacting and overlapping?  Gott's review of galaxy formation in 1977 \cite{Gott} described the unsettled state of affairs for the gravitational instability picture and Weinberg's text describes the difficulty of the early universe -- the hadron wall -- and a glimmer of hope that an ``elementary particle" description would offer.

During this period, cosmology was largely the province of astronomers and a few physicists (certainly less than 30 full time cosmologists worldwide).  The discovery of the CMB began to bring in physicists:  Penzias and Wilson (``radio astronomy" physicists), Wagoner and Fowler (who with Hoyle carried out the first modern calculation of BBN \cite{WFH}) and the Princeton gang who interpreted the Penzias and Wilson discovery (Dicke, Peebles, Roll, and Wilkinson \cite{Dickeetal}).  This foreshadowed the large number of physicist-cosmologists we see today, who would begin entering the field soon.

\subsubsection{Elsewhere}
I have focused on US astronomers and the influence of Weinberg's text, leaving out other cosmological activities.  Peebles wrote his first book, {\it Physical Cosmology}, around the same time as Weinberg, but it was much less influential amongst physicists, in part because Weinberg's text was so complete, beginning with a full development of General Relativity.\footnote{Further, his popular book, {\it The First Three Minutes}, published in 1977 had equations and was widely read by physicists.}   Peebles's second book, {\it Large-scale Structure of the Universe}, laid out a formalism for the study of cosmic structure and was influential amongst astronomers.

The controversy between the steady state theory and the big bang model was brief because the non-evolving steady state model was quickly ruled out by radio counts, quasars and finally the CMB \cite{Kragh}. Relativists including G.F.R. Ellis, Misner, Hawking, Penrose and the Russian school led by Zel'dovich and Novikov were looking for important applications of General Relativity and much of their work was on the nature on the initial singularity. 

Zel'dovich wrote two interesting reviews of cosmology itself -- one just before the discovery of the CMB \cite{Zel64} and one just after 
\cite{Zel67}.  The former argued against a hot beginning because it predicted a ``background radiation" of temperature 20\,K, which was inconsistent with observations.\footnote{While the details of his prediction were scant, the physics appears to be correct in contrast to the many predictions of Gamow and his collaborators \cite{Turner2021}.}  The latter exuberantly described all the new avenues of research that the discovery had opened up, as well as big questions that needed to be addressed.  

Finally, 
the observational side of cosmology was dominated by ``the Californians" with their big glass on Mt. Wilson (2.5\,m), Mt. Palomar (5\,m), and Mt. Hamilton (3\,m) for good reason.  Until 1973, 
the rest of the world did not have an optical telescope on a mountain top with aperture larger than 3\,m.\footnote{Because of its poor performance, I am not including the Russian 6-meter BTA-6 telescope. The US NSF built 4-m telescopes for public use on Kitt Peak (1973) and Cerro Tololo (1976); the 3.9-m Anglo-Australian Telescope in Siding Springs was built in 1974, and ESO built a 3.6-meter telescope on La Silla in 1977.}  The situation today is very different, with 19 optical telescopes of aperture larger than 6 meters around the globe, none in California.

\subsection{Technology}

Around 1970, astronomy was largely done with ground-based optical telescopes using photographic plates, which captured only 1\% of the incident light.  
Radio and microwave astronomy were still young, though both played a crucial role in falsifying the steady-state theory.  Computing did not yet play a decisive role in astronomy, though it was beginning to do so in other fields, notably high-energy physics.

The next fifty years would see epic changes in how astronomy is done:  the introduction of electronic detectors in the 1960s and 1970s (photoelectric devices), followed by CCDs in the 1980s, and the opening of new windows on the Universe, UV and IR in the 1970s, x-rays and gamma-rays in the 1970s and 1980s, neutrinos in the 2000s, and gravitational waves in 2016.  

The transformational role of NASA's  explorer program and its four ``Great Observatories" (Chandra, Compton, Hubble and Spitzer) cannot be overstated. In particular, the Hubble Space Telescope (HST) pinned down $H_0$ and opened our eyes to the high-redshift Universe, from supernovae to the birth of galaxies.  Its excellent, well-calibrated and reproducible observing conditions were necessary for and enabled precision cosmology.  


Today, gigapixel-CCD cameras collect almost 100\% of the incident light and enable photometry with 1\% precision; adaptive optics makes possible aperture-limited resolution on ground-based telescopes; and space-based observatories (US, European and Japanese) view the sky from the far IR to gamma rays.  Global facilities like ALMA provide unprecedented resolution and sensitivity for imaging distant galaxies.  And much more is ahead:  JWST, Euclid and the Roman Space Telescope in space, and the Rubin Observatory and two or three ELTs with adaptive optics on the ground.   

Amassing and analyzing the large, complex datasets that are being created is only possible because of the advances in computing and storage driven by Moore's Law.  Taking  advantage of the larger datasets being created and achieving precision cosmology requires complex simulations that connect the theoretical models with the raw data, taking account of how it is collected, from instrumental acceptance to seeing and foregrounds.  Astronomers pioneered the use of powerful Bayesian techniques that allow theories to be tested and compared, and multiple cosmological parameters to be extracted.  Things have come a long way from $\chi^2$ model testing!  

Advances elsewhere were crucial.  The search for dark-matter particles and proton decay led to the development of new instrumentation and the birth of ``underground astronomy."  Large ultra-sensitive, low-energy and low-background detectors are now deployed to detect dark matter and neutrinos in deep underground laboratories 
shielded from cosmic rays and other backgrounds.  Today, ``underground science" crosses multiple disciplines, from astronomy and physics to biology and geophysics.\footnote{Underground science was pioneered in 1965 by Ray Davis and his solar neutrino experiment in the Homestake mine.}


Martin Harwit's claim that progress in astronomy is driven by technological developments \cite{Harwit} is supported by cosmology.  CMB discoveries were enabled by ever increasing detector capability.  Photoelectric imagers enabled the discovery of flat rotation curves and the existence of dark matter.  
Wide-field CCD cameras and powerful computing made possible survey science possible and batch supernova discovery, which led to the discovery of cosmic acceleration.  

Finally, with technology came culture change.  50 years ago, an astronomer built the instrument, went to the mountaintop to observe, did the necessary theory, analyzed the data, and published the results, sending out a few hardcopy preprints after the paper was accepted.  Today, there are instrumentalists, theorists, observers, phenomenologists, and data specialists.  Queue observing and open-access archives (pioneered by HST) have redefined observing; collaborations have grown to 1000s, knit together by the web.  And even astronomers post their preprints on the arXiv before journal submission.   Without all of this change, many of the advances in cosmology would not have been possible.

\subsection{Origins of particle cosmology}

\subsubsection{$\alpha\beta\gamma$}
The roots of particle cosmology trace back to George Gamow and his famous 1948 $\alpha\beta\gamma$ paper, written with Alpher and Bethe, which introduced the idea of big-bang nucleosynthesis (BBN)  to explain the origin of the chemical elements \cite{abc}.  He and others realized that equilibrium nucleosynthesis in stars could not explain the abundance pattern of the chemical elements, and, in grand style, Gamow turned to the big bang for a wildly non-equilibrium event.  

Gamow et al. got much wrong, made several conceptually-flawed predictions for the temperature of the relic radiation -- the CMB -- and inadvertently invented the r-process (what their model basically was) \cite{MSTRF}.  Nonetheless, they got the big idea right:  nuclear physics in the radiation-dominated early phase of the Universe is important!  A year before the CMB was discovered, Hoyle and Tayler \cite{HoyleTayler} pointed out that BBN could explain the large primordial abundance of $^4$He. A year after the discovery, Peebles \cite{PJEPBBN} calculated how much $^4$He would be produced.  The first complete calculation, and the forerunner of all modern calculations, was that of Wagoner, Fowler and Hoyle \cite{WFH} in 1967.

Around this time, the landmark 1957 paper by Burbidge, Burbidge, Fowler and Hoyle \cite{BBFH}, which outlined the complex astrophysical story of how the elements in the periodic table are produced,\footnote{I have often heard that Hoyle initiated this program to keep pace with Gamow's model for the origin of the elements.  Others have reported that Hoyle disputed that claim \cite{PTletter}.} launched the field of nuclear astrophysics.  This vibrant field today encompasses astrophysical neutrinos, supernovae, and compact objects and has much overlap with cosmology.

In the 1970s, particle physics entered the picture with the Universe being deployed as a ``heavenly laboratory."  Three pioneering papers \cite{GZ,CMC,SM} 
used the relic abundance of neutrinos (about $113\,{\rm cm}^{-3}$ per species) and an estimate for the current mass density of the Universe to derive an upper bound to the neutrino mass.
Steigman, Schramm and Gunn used the BBN production of $^4$He to derive an upper limit to the number of light (mass $\ll\,$MeV) neutrino species (more light species leads to more $^4$He production) \cite{SSG}, and Lee and Weinberg used the mass density of the Universe to place a lower limit to the mass of a hypothetical heavy neutrino species (about 2\,GeV) \cite{LW, Dicusetal}.  

The hot big bang cosmology was being taken seriously enough to use the Universe to constrain particle physics.  What a turn around for the field where it was said that the errors are in the exponents!

I trace the birth of particle cosmology to 1979 and a series of papers about how the prevalence of matter over antimatter could have arisen in the early Universe \cite{Yoshimura, Weinberg, KTreview}. Now known as ``baryogenesis," its essential elements go back to a prescient paper written by Sakharov \cite{Sakharov}  in 1967.  

\subsubsection{Baryogenesis}
The idea is simple:  in the early Universe ($t \ll 1\,$sec), non-equilibrium interactions that violate $B$, $C$ and $CP$ allow the Universe to evolve a small, net baryon number.  When the temperature was much greater than the mass of the lightest baryon-number carrying particle species, baryons and antibaryons were as abundant as thermal photons, and the net baryon number corresponded to a slight excess of baryons over antibaryons.  When the temperature fell below the mass of the lightest baryon, essentially all of the antibaryons and most of the baryons annihilated, leaving the few baryons we see today for every 10 billion or so photons.\footnote{For many, until baryogenesis,  the puzzle was, how was the large entropy per baryon created? \cite{Zel67,highentropy,Misner}. Even at the center of a newly born neutron star, the entropy per baryon is only a few.  Some,  notably Penrose \cite{Penrose}, viewed an entropy of a billion per baryon as very small compared to what gravity could have created and wondered why we live in a low-entropy universe.}

\begin{figure}[h]
\includegraphics[width=0.9\textwidth]{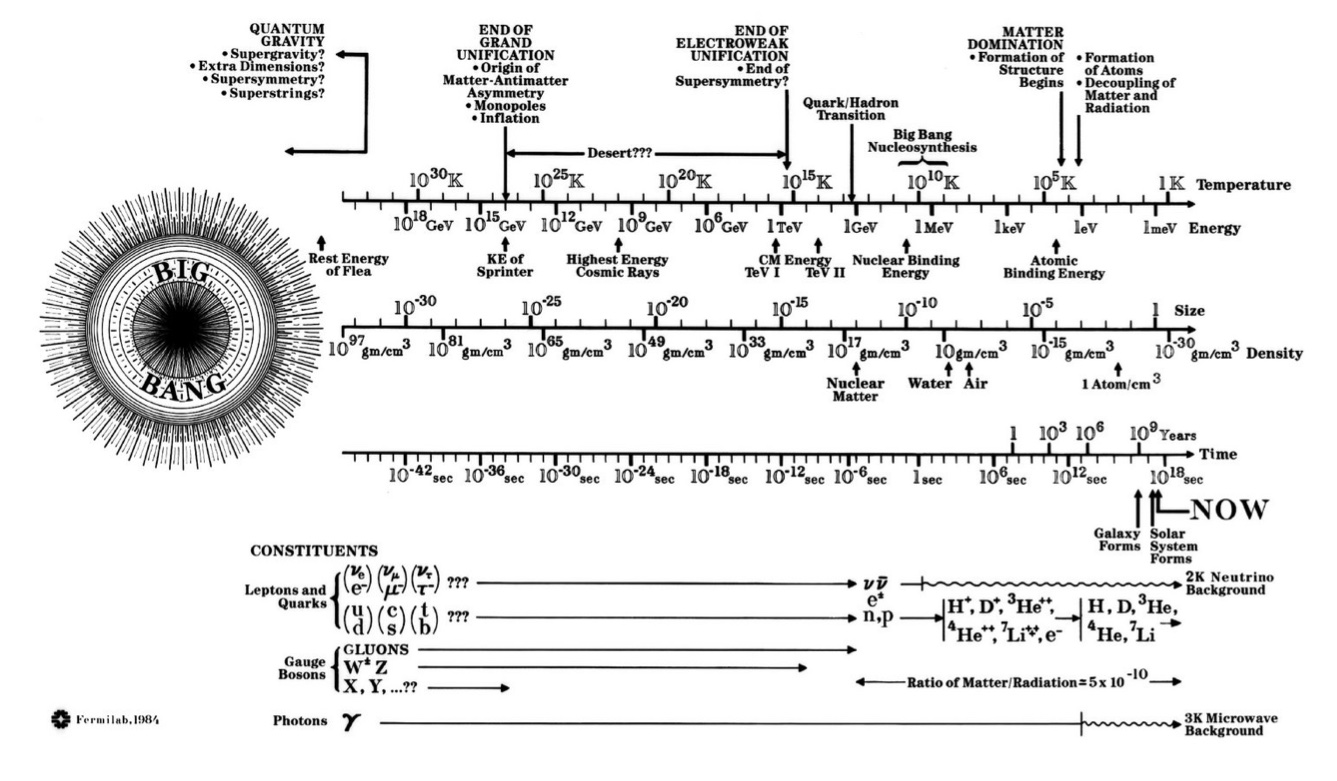}
\caption{Complete history of the Universe (a collaboration of Angela Gonzalez of Fermilab and the author, circa 1984).  This poster, which illustrates the inner space/outer space connection, became ubiquitous, gracing the walls of physics departments, DOE labs and even DOE headquarters.}
\label{fig:History}   
\end{figure}

Sakharov's idea seemed far out in 1967.\footnote{I have been told that when he first presented it in an April 1 seminar, his colleagues thought it was meant to be a joke.}  While $CP$ violation had recently been discovered, there was no reason to believe that baryon number was not conserved, and cosmological issues were not front and center for physicists.  By 1979, things had changed dramatically.  The discovery of asymptotic freedom \cite{GrossWilczek,Politzer} broke down Weinberg's hadron wall:  During the first microsecond, the Universe was comprised of a soup of weakly-interacting, point-like quarks, leptons, gauge and Higgs bosons of the Standard Model -- and probably other elementary particles.

Further, the convergence of the three coupling constants of $SU(3)\otimes SU(2)\otimes U(1)$ at an energy scale of $10^{15}\,$GeV or so led to grand unified theories.  Baryon-number violation was now mandatory, not crazy, and it had a laboratory signature: proton decay.  In 1980, the race was on to detect proton decay \cite{GUTs}.  Other cosmological consequences of GUTs followed: magnetic monopoles, phase transitions, inflation and particle dark matter.  The early Universe was open for business and the frontiers of particle physics and cosmology were converging. 

\subsubsection{Inner space and outer space meet at Fermilab}
My mentor, the late David Schramm \cite{DNSmemoir} was a powerful advocate for the convergence of the ``very small" and the ``very big," through his scientific contributions and leadership.  In 1982, he and then Director of Fermilab Leon Lederman challenged NASA to fund an astrophysics group at the Fermilab, to explore the then hypothetical deep connections between particle physics and cosmology \& astrophysics.  NASA did, through their innovative research program.

Lederman brought Edward (Rocky) Kolb and myself to Fermilab to lead it.  In May 1984, Fermilab hosted the coming out party for the new field, {\it Inner Space/Outer Space}.  More than 220 astronomers and physicists attended, and Steve Weinberg gave the summary talk.  The timing couldn't have been better, as the two big ideas -- inflation and CDM -- that were to underpin cosmology had just emerged.  To convince the scholarly UChicago Press to publish a mere proceedings, we boasted that it would mark the birth of a new discipline -- you can judge for yourself \cite{ISOS}.

The NASA Fermilab Astrophysics Center (NFAC) became the ``mother church" for this exciting interdisciplinary science and created the iconic poster linking the two fields seen on walls around the world (Fig.~\ref{fig:History}).  With a unique style, e.g., primordial pizza on Monday nights in my basement, topical workshops on cosmic strings, wormholes, and dark energy, all with Kolb-Turner designed conference T-shirts, NFAC trained many of the field's future researchers and leaders and sent them around the globe
 to spread the word.

NFAC led Fermilab to play a leading role in the Sloan Digital Sky Survey (SDSS) and to bring in many particle physicists looking for a new challenge.  SDSS achieved its primary scientific goal of determining the large-scale structure of the Universe to test inflation + CDM, but its larger legacy is the birth of survey science in mainstream astronomy.\footnote{Radio astronomers have long been doing surveys of the sky, in large measure because their antennae had large fields of view.}

Other big surveys followed in the footsteps of SDSS (and its sequels), including the Dark Energy Survey (DES), BOSS, DESI, LSST at the Rubin Observatory and soon the Nancy Roman Space Telescope.  This ``mega object" approach to astronomy dramatically changed the kind of science that could be done, the tools and skillsets needed to do it, and even the astronomy culture, including the size and composition of the collaborations.

The bold agenda of particle cosmology was laid out in the 2002 NRC report {\em Connecting Quarks with the Cosmos} \cite{Q2C}, and its 11 big questions were featured on the cover of {\em Discover} magazine \cite{Discover} as ``the greatest unanswered questions of physics."  Today, not only is particle cosmology front and center in most physics and astronomy departments, but also in all of DOE's high-energy physics labs.\footnote{Given all the pushback they received from DOE and the other Lab Directors about their venture, both Lederman and Schramm must be smiling somewhere.}

The particle cosmology agenda is featured in the priority-setting documents of both astronomy and particle physics
The April Meeting of the American Physical Society is now known as the $Q2C$ meeting, for Quarks to the Cosmos, and brings together 2000 or so astronomers, particle physicists, nuclear physicist and relativists.

While I have focussed on Fermilab's role in jumpstarting the field, important activities took place elsewhere, including workshops at the UCSB-ITP (including one of the very first ITP programs) and many summer and winter workshops at the Aspen Center for Physics.

\section{THE SECOND PARADIGM}
\subsection{Dark matter: a Kuhnian shift}

Today, dark matter is a central issue in  cosmology and astrophysics.  It wasn't always that way -- astronomy used to be the study of a Universe comprised of stars.  Dark matter fits the Kuhnian paradigm shift: evidence of an anomaly that is not paid attention to,  evidence grows in significance, and finally a shift to a new paradigm occurs.


As 
Morton Roberts recounts in his short history \cite{RobertsReview}, the evidence for something beyond stars began with Zwicky and the Coma cluster, where Zwicky \cite{Zwicky33} described the need for dark matter to hold the Coma galaxies together.\footnote{Dark matter was a small part of the paper which was more concerned with his tired light explanation of redshifts.  Oort, the year before, used the term dark matter to refer to the nearby disk matter that he inferred was there from stellar dynamics, but that wasn't seen \cite{Oort32}.  Oort's ``dark matter" is comprised of nearby faint stars, dust and gas clouds, and is not the dark matter that holds galaxies together.} Smith \cite{Smith36} showed the same was true for the Virgo cluster.  Babcock \cite{Babcock39} measured a rising rotation curve for M31 (Andromeda), and described it in terms of a rising mass-to-light ratio.  Even earlier, Lundmark \cite{Lundmark1930} found the same for M31 and a few other spirals.

Soon after neutral hydrogen (HI) gas clouds in the Milky Way were first detected with the 21-cm spin-flip transition \cite{Purcell1951}, radio astronomers began using 21\,cm to trace galactic rotation curves using HI,  probing them farther out than  optical measurements could at the time \cite{vandeHulst57,Bosma}.  By 1975, they had established that the rotation curve of M31 was not Keplerian \cite{Roberts1975}, but a connection to dark matter had not been made.  

The use of electronic detectors in place of photographic plates allowed Rubin and Ford to probe galactic rotation curves farther out than ever before, and in 1978 they published results from 10 high-luminosity spiral galaxies \cite{Rubin1978}.  All showed flat or rising rotation curves, with enclosed masses that grew linearly with galactocentric distance.  To quote from their concluding section, ``The observations presented here are thus a necessary but not sufficient condition for massive halos."  They also said, ``Roberts and his collaborators deserve credit for first calling attention to flat rotation curves."

In the early 1970s, theorists began to weigh in.  
Ostriker and Peebles asserted that spiral galaxies need massive haloes to stabilize their disks \cite{OstPee73}, and with Yahil \cite{OstPeeY74}, they made the case that with larger halo masses the estimates for $\Omega_0$\footnote{$\Omega_0$ is the mean mass density divided by the critical mass density, equal to one for a spatially flat universe.  Once the Universe something other than just stars, the different components -- stars, dark matter, baryons, and finally dark energy -- each got their own subscript: $\Omega_0 = \Sigma_i \Omega_i$.} should rise as well -- possibly to 0.2.  Einasto and his collaborators also made the case for massive galactic halos
connecting them to cluster dark matter as well \cite{Einasto1974}.  Reeves et al \cite{RAFS1973} argued that big-bang deuterium production, which falls rapidly with increasing baryon density, puts an upper limit on $\Omega_0$  of about 0.2.  

Around 1980, all the pieces were put together -- flat rotation curves, dark matter halos, rising mass-to-light ratios and cluster dark matter.  In my opinion, two papers mark the paradigm shift:  an influential review by Faber and Gallagher entitled, {\it Masses and mass-to-light ratios of galaxies} \cite{FG1979}, whose abstract's final line was, ``It is concluded that the case for invisible mass in the universe is very strong and becoming stronger," and Vera Rubin's {\it Science} article \cite{Rubin1980} summarizing her work on galactic rotation curves, whose abstract included, ``There is accumulating evidence that as much as 90 percent of the mass of the universe is nonluminous and is clumped, halo-like, around individual galaxies ... At present, the form of the dark matter is unknown."

In 1985, the IAU held its first symposium on the subject, {\it Dark matter in the Universe}, bringing together 190 researchers \cite{IAU117}.\footnote{In it, my 46-page (!) contribution explains and ranks the particle dark-matter candidates, top three: neutrinos, axions and photinos, puzzles about the ratio of particle dark matter to baryons and mentions the possibility of $\Lambda$.}  Dark matter was now officially a big deal.  However, there was still one big piece to come: dark matter is not made of atoms!

\subsubsection{Particle dark matter}
BBN deuterium production places an upper limit to the baryonic mass density, $\Omega_B< 0.2$.  In 1983, the use of all four light-element abundances identified a best-bet range, $0.015 \le \Omega_Bh^2 \le 0.026$ \cite{YTSSO}.  Provided the Hubble constant is greater than 50\,km/s/Mpc, baryons can at most contribution 10\% of the critical density, no matter what form they take.  This fact, the growing evidence for large amounts of dark matter -- perhaps even approaching the critical density --  the absence of a viable baryonic dark-matter candidate and the attractive candidates provided by particle physics opened the door for nonbaryonic, particle dark matter.  

First through the door were light neutrinos, where $\Omega_\nu h^2 = m_\nu/93\,$eV (per species).  A neutrino mass of around $25\,$eV and $H_0 =50\,$km/s/Mpc would give the critical-density Universe predicted by inflation.  In 1980, Schramm and Steigman's {\it A neutrino-dominated Universe} won the Gravity Research Foundation's annual essay competition \cite{SS1980}.
Adding wind to their sails, particle theorists had proposed the ``see-saw" mechanism that predicted eV neutrinos masses \cite{seesaw}, and, in 1980, a Russian experiment reported an electron-neutrino mass of $14\,{\rm eV} \le m_\nu \le 46\,{\rm eV}$ at 95\% confidence \cite{Lubimov1980}.  

A neutrino-dominated universe was short-lived.  The Russian experiment was a false alarm, and a neutrino-dominated universe
was ruled out because of disagreement between numerical simulations of how structure would form with the Universe we actually see \cite{DFW}.

More candidates came from particle physics.  Pagels and Primack \cite{PP} boldy proposed the first supersymmetric dark-matter candidate, a keV gravitino.  The axion followed \cite{PWW,IpserSikivie}; and it ushered in the cold dark matter (CDM) scenario \cite{TWZ}.  Next came the neutralino \cite{neutralino}, and it has attracted the most attention since.

\paragraph{Neutralino}
In most supersymmetric (SUSY) extensions of the Standard Model the lightest SUSY particle is neutral, stable and has a mass expected to be between $100\,$GeV and a few\,TeV.  Called the neutralino, it is a linear combination of the Zino, photino and higgsino, the SUSY partners of the Z boson, the photon and the Higgs.  Further, its interactions with ordinary matter are weak, which means neutralinos can be produced at an accelerator and that halo neutralinos can be detected, and it behaves like cold dark matter.  It is an excellent dark matter candidate, but its path to front runner was complicated. 

It began with a false start, the gravitino, the SUSY partner of the graviton \cite{PP}.  In the Pagels-Primack model, the lightest SUSY particle is the gravitino, not the neutralino.  Further, the gravitino behaves like warm dark matter, not cold dark matter \cite{BondSzalayTurner}.  When the neutralino arrived, theorists were at first timid, writing ``limit-setting" papers, not neutralino dark-matter papers \cite{Ellis83,Ellis84,Goldberg}.  By 1984, it was all sorted out and the Inflation + CDM paradigm was in place, with the neutralino as the leading CDM candidate.

The neutralino's attractiveness as a dark matter candidate goes  beyond the appeal of SUSY in particle physics and its detectability. There is another pull, often referred to as  {\it the WIMP Miracle}, a term I hate.  I would prefer, ``the WIMP hint?"  

The neutralino's presence today as dark matter owes to the ``freezing out" of its annihilations in the early Universe and the subsequent ``freezing in" of its abundance.  When the temperature of the Universe was much greater than its mass, neutralinos were in thermal equilibrium with an abundance comparable to photons; as the temperature fell below its mass, neutralinos should have annihilated to exponentially small numbers.  However, as they became less abundant, they ceased annihilating, leading to freeze out.

Calculations show that the relic mass density scales inversely with annihilation cross section and is comparable to the critical density for a weak-interaction annihilation cross section, $\sigma \sim 10^{-36}\,{\rm cm}^2$ \cite{KolbTurner}, like that of a neutralino, or any particle whose annihilation cross section is weak.  (WIMP, for weakly-interacting massive particle, is the term I coined for such particles \cite{WIMP}.)
Whether or not this fact is a hint or a misleading coincidence remains to be seen.\footnote{Shortly after the discovery of the CMB, Hoyle and his colleagues pointed out that the energy released by an early generation of stars that made the large ``primordial" $^4$He abundance would be comparable to that in the CMB and could explain the CMB if it could be thermalized \cite{WFH}.  It can't be thermalized and so this hint was just misdirection.}

\paragraph{Desperately-seeking dark matter} Today, the $100\sigma$ discrepancy between $\Omega_Mh^2$ and $\Omega_Bh^2$ (see \S4.1) makes an airtight case for non-baryonic dark matter.  Namely, there is much more matter than can be explained by baryons alone.  But a $100\sigma$ discrepancy doesn't make particle dark matter a fact -- it could just be evidence for a coming paradigm shift.  Seeking additional, independent evidence and identifying the dark matter particle has become a holy grail of cosmology, particle physics and astrophysics.

Neutralinos interact weakly with ordinary matter, opening three avenues for additional evidence:  detection of the kinetic energy halo neutralinos deposit by scattering off nuclei in a detector \cite{GoodmanWitten}, their creation at particle accelerators, and detection of annihilation products from halo neutralinos or those captured by and annihilating in Earth or the sun.  All three methods now have the sensitivity to probe interesting regions -- but not all -- of the neutralino parameter space; none has produced a definitive signal \cite{DMReview}.  These searches also apply to other WIMPs, because the key is weak interactions with ordinary matter.

Much effort has gone into searching for axions, based upon a proposal by Sikivie \cite{Sikivie1983}:   the conversion of halo axions traversing a very strong magnetic field  into microwave photons.  The field has to be very strong and a high-Q cavity is needed to collect and detect the photons.  The necessary sensitivity has been achieved, but only a small fraction of the most promising mass range ($10^{-6}\,$eV to $10^{-3}\,$eV) has been probed, and without success \cite{ADMX}.

Finding independent evidence for the dark matter particle would not only complete the dark matter story, but also it would open a new window on early Universe cosmology, akin to BBN.  However,  all we know for certain is that dark matter particles have gravitational interactions, move slowly (i.e., are cold) and don't interact like baryons.  They could have no interactions with baryons, or very, very weak interactions with them making dark matter particles undetectable, in spite of their great abundance and impact.

\subsubsection{False starts}
In 1983, Milgrom noticed that the need for dark matter in galaxies occurs not at a particular distance from the center of the galaxy, but at an acceleration,  $a_0 \simeq 10^{-8}\,$cm/sec$^2$, and proposed MOND for MOdified Newtonian Dynamics \cite{MOND}.  MOND posits that $F \simeq ma^2/(a+ a_0)$, so that for $a\ll a_0$, $F \propto a^2$.  This one-parameter model fits all the rotation-curve data, but can't account for cluster dark matter and makes no other predictions. In short, it can't be falsified.  Nonetheless, the fact that the need for dark matter in galaxies occurs at an acceleration $a_0$ must be explained \cite{KapTurner}.


Before the argument against baryonic dark matter was airtight, dark baryons in the form of optically faint objects (e.g., brown dwarfs, white dwarfs, black holes or Jupiters) were still a possibility, but hard to test.  In 1986, Paczynski \cite{Paczynski} proposed using gravitational microlensing, the  brightening of a star as a object passes along its line of sight to us, to detect halo dark matter in the form of MACHOs, for Massive Astrophysical Compact Halo Objects.\footnote{Coined by Kim Griest, in response to WIMPs.} In 1993, the first microlensing events were announced  \cite{MACHO1}, and for a while, as more events  were seen, it looked as if the Milky Way might have a MACHO halo.  However, many of the events turned out to be other things that went ``bump in the night," and the current consensus is MACHOs do not make up our halo \cite{MACHO3}.  

Quark nuggets was the most intriguing suggestion.  Witten \cite{QNWitten} pointed out that if the quark/hadron transition were first order and if objects with large baryon number were more stable than nucleonic matter, a majority of the baryons could end up in quark nuggets rather than nucleons after the quark/hadron transition.  This could explain the 5-to-1 dark matter-to-baryonic matter ratio and get around the BBN bound without the new need for a new form of matter.  However, a better understanding of QCD has ruled this out.

I close by mentioning two  comprehensive reviews of the dark matter story, the first from a particle physicist's perspective \cite{BertoneHooper2018}, and the second from an astronomer's perspective \cite{Trimble}.  And by pointing out that dark matter has become such a cultural touchstone that even biologists have borrowed it, using it  as an informal term for poorly understood genetic material,  unclassified microorganisms and ``junk" DNA in the genome.  

\subsection{The CMB and the birth of precision cosmology}
It would be hard to overstate the role of the CMB in ushering in precision cosmology, and it is still its best exemplar.   Why?  First, the underlying physics is simple:  atomic and gravitational physics and linear perturbations of the homogeneous and isotropic background cosmological model.  Next, the relationship between temperature variations and the underlying density perturbations that create them is straightforward: the response of the baryon-photon fluid to the dark-matter inhomogeneities, which depends upon the baryon and dark matter densities, the expansion rate, the atomic composition and the geometry of the Universe. This means that the anisotropy of the CMB can be used to reveal much about the Universe.  Finally, the individuals involved in making CMB measurements were (and still are) largely physicists, who brought their rigorous laboratory approach.

Because CMB anisotropy and its polarization have yielded so much precision information about the Universe, I mark the COBE satellite's detection of CMB anisotropy in 1992 \cite{COBEDMR}  as the official beginning of precision cosmology, 

\subsubsection{The long road to COBE}
In their discovery paper, Penzias and Wilson noted that the CMB was isotropic at the 10\% level \cite{PW}.  Peebles and others realized that if structure formed by gravitational amplification of small density inhomogeneities \cite{Lifshitz}, they should leave their imprint in the anisotropy \cite{SW,deltaToverT}.  Anisotropic expansion or the rotation of the Universe would give rise to a quadrupole anisotropy.  Rees showed that the CMB anisotropy should polarized at the few percent level \cite{ReesPol}.  And Sunyaev and Zel'dovich pointed out that spectral distortions would arise in the CMB if the line of sight passed through clusters and scattered off the hot gas in them on the way to us \cite{SZ}.

While CMB anisotropy had to be present and was clearly important, well-founded predictions for its level were lacking.  Nonetheless, a generation of talented experimentalists continued to place more and more stringent limits on CMB anisotropy at various scales, from the dipole and quadrupole to the arcminute angular scales.  

In 1977, a definitive detection of the dipole anisotropy was made from data collected by a re-configured U2 spy plane \cite{Smootetal}.  The dipole arises due to our motion today with respect to the cosmic rest frame and not the underlying density perturbations at last scattering.  The magnitude and direction of the ``kinematic dipole" is known with great precision.  The solar system moves with at a speed of $369.82 \pm 0.11\,$km/s in the general direction of the constellation Leo ($l = 264.021^\circ \pm 0.011^\circ$ and $b=48.253^\circ \pm 0.005^\circ$) \cite{PlanckLegacy}.

By 1980, limits to the CMB anisotropy on various angular scales were:  $\delta T/T < 10^{-4}$ \cite{DTWRoySocReview}.
Theorists were getting more serious about their predictions, e.g., including non-baryonic dark matter \cite{WilsonSilk}.  In 1984 the first calculations of CMB anisotropy in CDM models were done \cite{BE1,VittorioSilk}, and today's ubiquitous angular power spectrum was introduced \cite{BE2}.  The CDM prediction, $\delta T/T \simeq 10^{-5}$, was a factor of ten below the limits.

\subsubsection{COBE}
In 1976, NASA approved the Cosmic Background Explorer (COBE) mission, a merger of three concepts (DMR, FIRAS and DIRBE).  It was designed to fly on the Space Shuttle, but the Challenger accident changed that, and COBE was launched on 18 November 1989 from Vandenberg AFB  on a Delta launch vehicle.  

\begin{figure}[h]
\begin{subfigure}{0.45\textwidth}
\includegraphics[width=0.95\linewidth, height = 3.6cm]{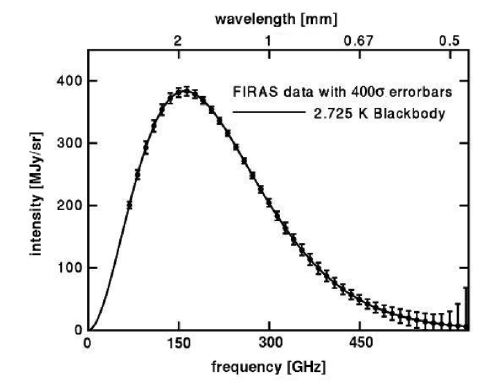}
\end{subfigure}
\begin{subfigure}{0.55\textwidth}
\includegraphics[width=0.95\linewidth, height = 3.5cm]{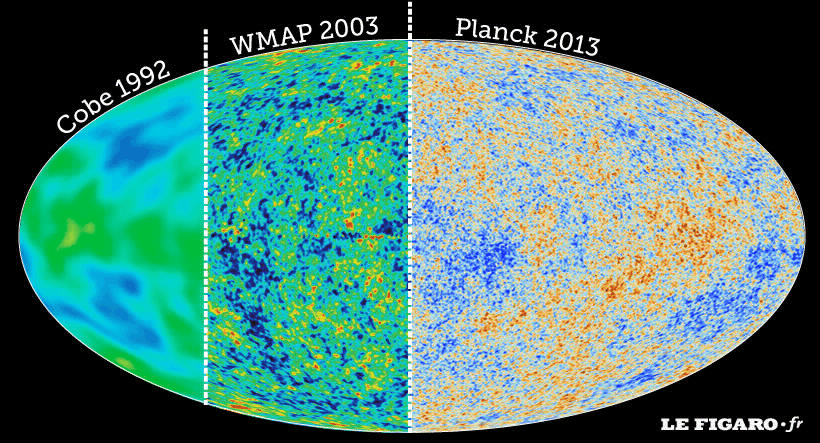}
\end{subfigure}
\caption{Birth of precision cosmology. left: The ``perfect blackbody" spectrum measured by the FIRAS instrument on COBE.  right: The CMB anisotropy progression, with increasing sensitivity and resolution, from COBE to WMAP to Planck.}
\label{fig:COBE}
\end{figure}

Less than two months later, John Mather showed the first  spectrum of the CMB produced by his FIRAS instrument on the last day of the January 1990 AAS meeting in Washington, DC.  It was a perfect blackbody over the range 60 to 640\,GHz (see Fig.~\ref{fig:COBE}), and he received a standing ovation.   Ultimately, FIRAS pinned down the CMB temperature to 4 significant figures, $T = 2.7255 \pm 0.00057\,$K \cite{FIRAS}. Precision cosmology, indeed.


The DMR team would make its big announcement two years later, at the April 1992 APS meeting, also in Washington, DC and also on the final day.  They announced the first detection of CMB anisotropy beyond the dipole, at a level of $\delta T/T \simeq 10^{-5}$ on the angular scale of $10^\circ$.\footnote{The DMR did  not have much margin to spare: the year 1 data had a S/N of about 1 per pixel (roughly 100 sq. deg. patch) which increased to about 2 in the final 4-year dataset.  The detection of anisotropy on the $10^\circ$ scale was highly significant, but the features in the map much less so \cite{DTWPNAS}.}  Stephen Hawking called it, ``the greatest discovery of all time."  Overstatement to be sure, but COBE had opened the era of precision cosmology.

Inflation set the shape of the spectrum of density inhomogeneities (scale-invariant), and so COBE could be used to set the overall amplitude.  In turn, this allowed comparison with measurements of the inhomogeneity today from the distribution of galaxies.  

With its $10^\circ$ beam, COBE had only probed ``the Sachs-Wolfe plateau" \cite{SW} and not the acoustic peaks, where all the information about cosmological parameters lie.  
In 1992, the race was on to design and build experiments with greater angular resolution and sensitivity to read cosmology's Rosetta Stone \cite{PhysicsToday}.  

\subsubsection{After COBE}
Two satellite missions were approved in 1996, NASA's WMAP,\footnote{David Wilkinson, who played a leading role in the design of the Microwave Anisotropy Probe (MAP), died in September 2002, and in February 2003, MAP became WMAP in his honor.} which launched in 2001 and announced its first results in 2003, and ESA's Planck, which launched in 2009 and announced its first results in 2013.  In the meantime, a host of ground-based and balloon-borne experiments were mounted, where new technology, e.g., bolometers,  and new techniques (use of the angular power spectra) could be rapidly deployed.  

Theorists were busy too, emphasizing the treasure trove of information contained in the acoustic peaks and ``forecasting" how well they could be measured, often influencing the designs of the experiments \cite{Knox,Kamio1}.  The big prize was the curvature of the Universe, encoded in the position of the first peak, with the flat-Universe prediction of inflation corresponding to $l\simeq 200$ or about $1^\circ$.  

They also pointed out that by analyzing the polarization of the CMB anisotropy in terms of the even (E) and odd (B) parity modes, one could hunt for the signature of the gravitational waves produced by inflation:  only the gravitational waves excite the B-mode \cite{Kamio2,Seljak}.  (The story is of course more complicated as gravitational lensing by intervening large-scale structure distorts the CMB anisotropy and can transmute the E-mode polarization produced by density perturbations into B-mode polarization \cite{WHu}).

\begin{figure}[h]
\begin{subfigure}{0.5\textwidth}
\includegraphics[width=0.95\linewidth, height = 4.5cm]{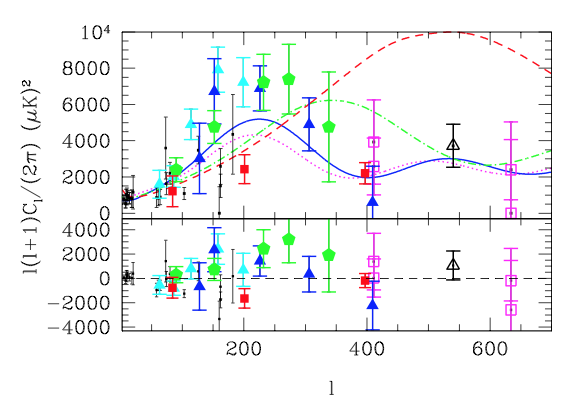}
\end{subfigure}
\begin{subfigure}{0.5\textwidth}
\includegraphics[width=0.95\linewidth, height = 4cm]{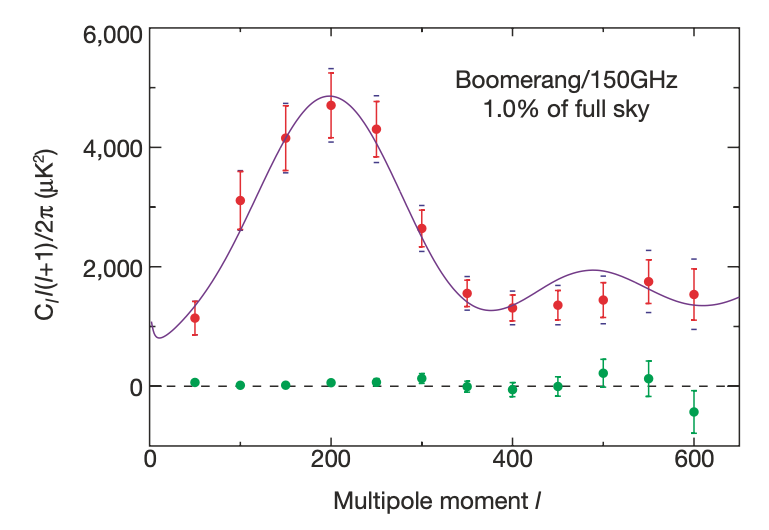}
\end{subfigure}
\caption{The first peak. left: Circa 2000, emerging from COBE (black) and ground-based experiments (different symbols); solid curve is a flat Universe, broken curves are open models \cite{DodelsonKnox}.  right:  Boomerang angular power spectrum clearly shows the first peak at $\ell = 200$, as expected for a flat Universe with $\Omega_B = 0.05$, $\Omega_M = 0.3$, $\Omega_\Lambda = 0.7$ and $h=0.70$.}
\label{fig:Boom}
\end{figure}


\paragraph{Boomerang} There was steady progress with results from a number of different experiments filling in the angular power spectrum and the first acoustic peak was beginning to appear in combined data at $l\sim 200$ \cite{DTWPNAS}; see Fig.~\ref{fig:Boom}. However, the shot heard 'round the world came in April 2000, with the announcement of results from Boomerang, an Antarctic balloon experiment led by Andrew Lange\footnote{Lange died tragically in 2010; his CMB legacy continued with bolometric detectors his group designed being deployed on Planck and the Keck/BICEP array.} \cite{Boomerang2000}.  

Their power spectra, from $l=50$ to $l=600$ and at four frequencies, revealed a clear peak at $l = 200$ and a good fit to today's consensus $\Lambda$CDM model (see Fig.~\ref{fig:Boom}).  At 95\% confidence, $\Omega_0$ was between $0.88$ and $1.12$.  Boomerang's result firmly established $\Lambda$CDM.


Much more would follow:  the first detection of CMB polarization by DASI in 2002 \cite{DASI}, the definitive detection of the SZ effect and discovery of clusters using the SZ signal alone \cite{SZ}, the ``TE" and ``EE" angular power spectrum, the detection of B-mode polarization from both dust emission and gravitational lensing, and limits to gravitational-wave produced part \cite{Bmode}.  

Today, the combination of WMAP and Planck and two ground-based CMB experiments (SPT and ACT) have  measured the anisotropy of the CMB with cosmic-variance precision from $l=2$ to $10,000$ and the hunt is on for the B-mode polarization produced by inflationary gravitational waves.  

For 10 years, the Keck/BICEP experiment at the South Pole has led the hunt and has the best upper limit: $r < 0.04$ at 95\% confidence \cite{Bmode}.  That is, gravitational-wave produced anisotropy is at most 4\% of that produced by density inhomogeneities.  The CMB Stage IV (CMB-S4) experiment, which is being developed and involves 21 telescopes at the South Pole and in the Chilean Atacama desert, aims for a sensitivity of $r = 10^{-3}$ \cite{CMBS4}.  

There is no consensus prediction for $r$, with some theorists arguing that it {\it must be larger} than $0.01$ \cite{larger}, others saying they can prove that it {\it must be much, much smaller} than $10^{-3}$ \cite{Lyth}, and models of inflation predicting all numbers in between.  All that can be said with certainty is detecting the signature of these gravitational waves would reveal when inflation took place and teach us about the underlying physics \cite{TurnerGW}.

\subsection{Big ideas: inflation + cold dark matter}
While 1980 marked the birth of particle cosmology, it wasn't until 1984 that inflation + CDM arrived and became a guiding light for cosmology.  It
built upon and dramatically expanded the very successful hot big bang model.  Its bold and falsifiable predictions caught the attention of those in the the field and attracted both theorists and experimentalists from outside the field, many from particle physics.  


\subsubsection{Inflation}
The story begins with superheavy magnetic monopoles, which in addition to baryon number violation and neutrino mass, are predicted by Grand Unified Theories
\cite{PreskillMonopoleReview}.  Moreover, they should be produced in the early Universe during the phase transition where the GUT symmetry is spontaneously broken to that of the Standard Model.  And further, 
a Stanford low-temperature physicist had detected a signal consistent with a magnetic monopole traversing a superconducting loop is his lab \cite{Cabrera}.  

There was only one problem:  a calculation by Harvard graduate student John Preskill \cite{Preskill} clearly showed that magnetic monopoles should have been produced in numbers far too large to be consistent with the hot big bang cosmology.

There was a possible out:  dilution of the monopoles by the entropy produced in a first-order GUT phase transition.  In the absence of entropy production, the number of photons in the Universe is conserved.\footnote{More precisely, the number of relativistic particles; see Chapter 3.4 of \cite{KolbTurner}.}  The number of monopoles here today is simply the monopole-to-photon ratio produced in the early Universe times the number density of photons today (about $411\,{\rm cm}^{-3}$).  Thus, the monopole problem can be solved by the creation of additional photons, which reduces the monopole-to-photon ratio.

Many pursued this fix. Guth's paper \cite{Guth} stood out for its expansiveness and clarity.  He showed that not only does a strongly-first-order phase transition produce entropy, but also, when the Universe is stuck in the old phase, ``false-vacuum energy" behaves like a cosmological constant and the Universe expands exponentially. Moreover, the exponential growth and entropy production could solve two other cosmological problems, the horizon and flatness problems, which were less well appreciated, especially by particle physicists.  And finally, a phase transition that takes long enough to solve all three problems, never finishes and leaves a Universe that is a mess, i.e., no ``graceful exit." Got it, a really good idea that doesn't work.

A word about the horizon and flatness problems \cite{Dicke}.  While the Universe was much smaller earlier on, it was also expanding rapidly and the distance that a photon could have travelled since the beginning  until a given time -- called the distance to the horizon -- is not infinite, but rather a factor of order unity times the age of the Universe at that time. The CMB is very smooth; unless the Universe began smooth, causal interactions could only smooth it up to scales of about $1^\circ$, the size of the horizon at last scattering as seen on the CMB sky today.  The horizon problem also prevents any causal process from creating the small amount of lumpiness needed to seed structure in a smooth Universe.  The flatness problem involves the fact that with time, the deviation of $\Omega$ from 1 grows; unless $\Omega$ were initially very, very close to 1, the Universe would have long ago collapsed or gone into to free expansion, leading to $\Omega_0 \ll 1$, in conflict with measurements. 

The horizon and flatness problems were part of the larger issue of the  special initial conditions required to achieve a Universe similar to ours -- long-lived, smooth and flat, which relativists \cite{Relativists,Misner} had been worrying about long before Dicke \cite{Dicke}.  To address these issues, Penrose was advocating for rules about the kind of initial singularities permitted, and has advocated his Weyl curvature hypothesis as an alternative to inflation \cite{Penrose}.

Back to the main story, Guth made his case for inflation well.  People paid attention, and within about a year, a solution to the graceful-exit problem was put forward by Linde \cite{Linde} and by Albrecht \& Steinhardt \cite{AS,ASTW}:  ``slow-roll inflation."

\subsubsection{Nuffield} 
\begin{figure}[h]
\begin{subfigure}{0.3\textwidth}
\includegraphics[width=0.95\linewidth, height = 5.2cm]{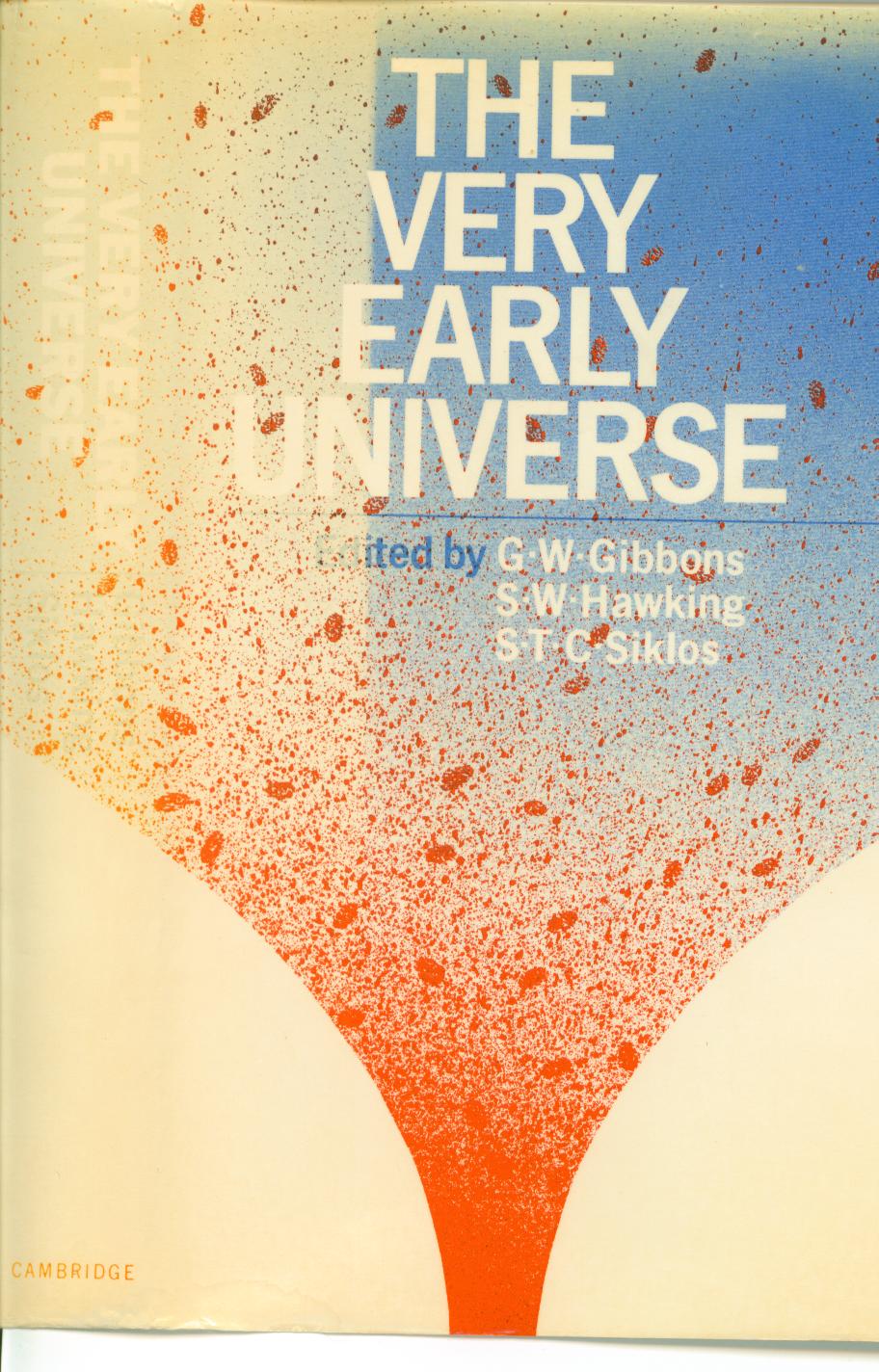}
\end{subfigure}
\begin{subfigure}{0.70\textwidth}
\includegraphics[width=0.95\linewidth, height = 5cm]{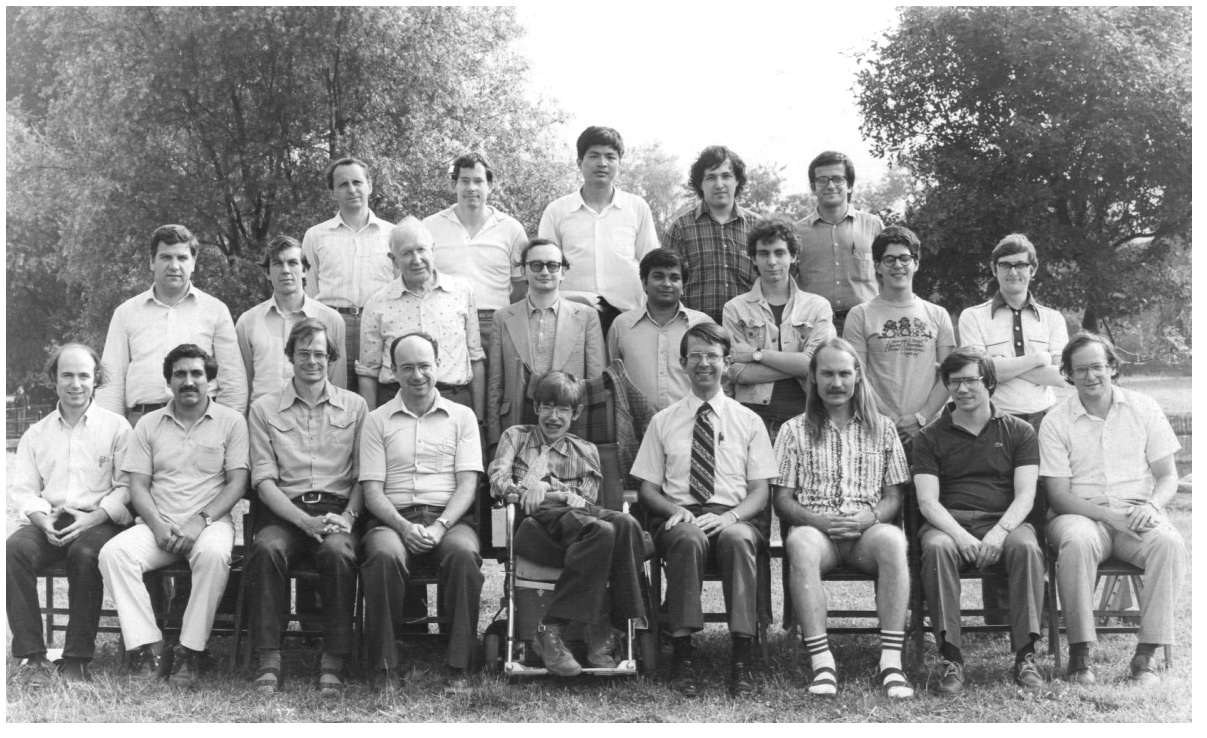}
\end{subfigure}
\caption{The 1982 Nuffield Workshop.  left:  the cover of the proceedings.  right:  F. Wilczek (front row, far left); S. Hawking (front row, center); J. Bardeen (left of Hawking); the author, A. Guth and P. Steinhardt (front row, far right); A.A. Starobinsky (above Hawking).  J. Preskill, above and to the left of Starobinsky. A. Linde is not in the photo.}
\label{fig:Nuffield}
\end{figure}

All the pieces came together at the Nuffield Workshop on the Very Early Universe, organized by Stephen Hawking and Gary Gibbons at Cambridge University, in the summer of 1982 (see Fig.~\ref{fig:Nuffield}).  The details of ``slow-roll" inflation and the quantum origin of density fluctuations were thrashed out in real time \cite{TheVeryEarlyUniverse,BST,GuthPi,Starobinskii,Hawking}, and I can say that Nuffield was the most exciting meeting of my scientific career.

While the original Linde/Albrecht/Steinhardt model didn't work because it resulted in density perturbations that were too large, by the end of the workshop there was a ``prescription" for successful inflation \cite{prescription}, and new models began springing up.  Today, there is no shortage of viable models of inflation, but also no standard model.

The same basic idea still underlies almost all models of inflation today.  A scalar field  rolls slowly down a very flat potential having started far from the minimum.  While it slowly rolls, its potential energy behaves like a cosmological constant and drives a nearly exponential expansion, taking a tiny bit of the Universe that could have been smooth and blowing it up to enormous size.  The wavelengths of quantum fluctuations in both the scalar field and in the spacetime metric are similarly blown up, to astrophysical size.

Inflation ends when the scalar field reaches the bottom of its potential, and that potential energy is converted into particles, thereby reheating the Universe and creating an enormous amount of entropy.  The quantum fluctuations become density perturbations and gravitational waves respectively, and the Universe begins its usual quark soup phase.

By 1983, inflation had become the driving  force in cosmology because of the power of its three big predictions:  (i) flat Universe ($\Omega_0 = 1$); (ii) almost scale-invariant spectrum of nearly-Gaussian density (curvature) perturbations;\footnote{The scale-invariant spectrum of perturbations was first proposed by Harrison \cite{Harrison}; its special feature is that, unlike any other spectrum, it does not diverge on either small or large scales and does not need to be cutoff. This virtue was well-appreciated by cosmologists.} and (iii) almost scale-invariant spectrum of gravitational waves \cite{GWs}.  

The first prediction made inflation both bold and falsifiable:  the observational evidence at the time was $\Omega_0 \sim 0.1$, and so there must be something in the Universe in addition to baryons.
The second prediction fixed the shape of the spectrum of density perturbations, which could then be normalized by using the level of inhomogeneity on the scale $8h^{-1}\,$Mpc today (about unity).  This set the overall amplitude at the level of $\delta \rho /\rho \sim 10^{-5}$.  This is the hardest constraint for inflation model-builders and leads to the necessity of a very flat scalar-field potential.  

Together, the first two predictions provide the initial conditions for the formation of structure,
and growing computing power made possible more and more realistic simulations of structure formation.  The results could then be compared to the growing body of data about the large-scale structure of the Universe, leading to a renaissance in the study of structure formation as well as a key test of inflation.

The third prediction is the most intriguing.  While, the amplitude of density perturbations needed to explain cosmic structure was known, there was no such prediction for gravitational waves.  Gravitational waves of any kind had yet to be detected and the challenge of doing so was daunting.\footnote{The decay of the orbit of the binary pulsar provided strong indirect evidence for their existence \cite{HulseTaylor}, but the LIGO, which would detect them in 2016, would not begin for ten years.}  

Detecting these gravitational waves would reveal when inflation took place: their dimensionless strain amplitude is simply the expansion rate during inflation divided by the Planck mass,  $h\sim H/m_{\rm pl}$.   And, a few years later, a consistency relationship between their amplitude and the deviation of their spectrum from scale-invariance was discovered \cite{consistency} as well as ideas about how to reconstruct the scalar field potential \cite{Lidseyetal}.  

Last but not least, it could be -- and was -- said that a flat Universe and scale-invariant perturbations were the attributes of any sensible cosmology, and thus not a strong test of inflation.  The same could not be said of gravitational waves or the slight deviation of both density perturbations (and gravitational waves) from scale-invariance.  While not appreciated early on, the prediction of not quite scale-invariant density perturbations has become a key test of inflation, which it passes with flying colors (see \S4.1.1).   And as described in \S3.2.2.1, the most powerful way of detecting gravitational waves is through the B-mode imprint they leave on the CMB polarization.   Inflation-produced gravitational waves remain as a big prize and test of inflation.

\subsubsection{Cold Dark Matter (CDM)} 
Inflation needed something beyond baryons to reach the prediction of $\Omega_0=1$.
Particle dark matter provided the most compelling possibility.  There are two limiting cases:  hot dark matter (HDM) where the DM particles move fast and stream out of the small perturbations that make galaxies and clusters, and cold dark matter (CDM) where the DM particles move slowly and don't stream out.\footnote{There is an intermediate case, warm dark matter where the streaming scale is the size of galaxies \cite{BondSzalayTurner}.}

With HDM structure forms from the ``top down," very large structures form first and fragment  to form smaller objects; HDM was ruled out almost immediately \cite{DFW}. CDM, on the other hand, where structure forms from the ``bottom up," galaxies, followed by clusters and superclusters, looked much better.  Its broad-brush predictions were fleshed out very quickly \cite{blumenthaletal,Bardeenetal}, and by the time of the {\it Inner Space/Outer Space} meeting in mid-1984, Inflation + CDM was firmly in place; see \cite{ISOS}.

\subsubsection{Defective defects} Revolutions aren't linear, even in cosmology. There was another worthy idea that aspired to lead cosmology forward, a network of one-dimensional topological defects produced in a GUT-scale phase transition, cosmic strings \cite{CS1}.  Such a network should evolve as strings collide and intercommute to produce a self-similar distribution of string loops that act as seeds for structure formation.  Further, the oscillations of string loops produces a spectrum of gravitational waves that might be detected and long strands of string leave a unique signature in the CMB. 

For a time, strings were ``Pepsi" to inflation's ``Coke." The end for cosmic strings came when the first acoustic peak was detected \cite{Boomerang2000}, because they predicted no acoustic peaks \cite{CS2}.  Like a strong competitor often does, cosmic strings made inflation's triumph look all the more impressive.  Topological defects are so attractive that there is continued interest in them, with the hope that some remain from early on and can be detected \cite{CS3}.

\subsection{Cosmic acceleration and dark energy}
It all came together in 1998.  The two supernova teams announced their discovery of cosmic acceleration early in the year \cite{SCP,hiz}, by mid-year $\Lambda$CDM was the new cosmological paradigm, and the year ended with {\it Science} magazine naming cosmic acceleration ``the breakthrough discovery of 1998" \cite{ScienceDec1998}.  Cosmic acceleration was called one of the biggest surprises of all time; in fact, the long lead up to it, may have made it the most anticipated one.

\subsubsection{Troubles with CDM}
The story begins in 1984.
Numerical simulations had quickly dispensed with HDM, and similar simulations of CDM reproduced the large-scale structure revealed by the CfA1 survey \cite{EWD}, making Inflation + CDM look like a winner.  However, there were dark clouds on the horizon.

First, the observational data suggested that $\Omega_0 \sim 0.1$, and not close to 1 \cite{FG1979}; second,
CDM predicted a highly-evolving Universe for which there was no evidence (and conventional wisdom, loudly espoused by Peebles, that galaxies were in place at redshift $z\gg 10$); and third, to get the large-scale structure just right, one needed  $\Omega_M h \sim 0.2 - 0.3$.

None of these problems was a show stopper (yet). After all, the full extent of galactic halos was not yet known, and estimates of $\Omega_M$ were still increasing as the scale of the region probed  increased  \cite{FG1979}.   Further, few observations probed the Universe beyond redshifts of a few tenths, and thus could not rule out lots of evolution at modest redshifts.  In fact, there was evidence for such evolution, the so-called Butcher-Oemler effect, where more blue (star forming) galaxies were observed in clusters at redshifts $z\sim 0.3$ than at lower redshift \cite{BO}. 

The third problem could be dismissed by pointing out that numerical simulations were not yet very sophisticated and the CfA1 galaxy survey only had 2400 galaxies.  Shifting ``$\Omega_M h = 0.2 -0.3$" to $=0.5$ or so, might be just fine since the age of a flat Universe, $2H_0^{-1}/3 \simeq 7h^{-1}\,$Gyr needed a small Hubble constant to accommodate the oldest stars whose ages were estimated to be 10 to 15\,Gyr.

\subsubsection{$\Lambda$ again}
The ``$\Omega$-problem" seemed the most challenging; namely, how to make a flat Universe consistent with data that indicated $\Omega_0 \sim 0.1$ \cite{TSK}.  Remedies were suggested \cite{TSK,PJEPLambda,GELambda}, including the possibility of a cosmological constant. 

$\Lambda$ had much to recommend it:  it was a smooth component of energy density that would escape detection and didn't impede the growth of structure significantly. It  ``lengthened" the age of Universe, for $\Omega_M \sim 0.3$, the age $t_0 \sim H_0^{-1}$; and it was highly testable. 

But, $\Lambda$ had a  checkered history too, having been invoked for every crisis du jour in cosmology \cite{DEARAA,EPJH}:  by Einstein to get a static Universe, by Eddington to resolve the age crisis that arose due to Hubble's very large ``Hubble constant" ($H_0 \sim 500\,$km/s/Mpc), by Hoyle and Bondi \& Gold in their steady state model, and in the 1960s to explain the preponderance of quasars at redshifts $z\sim 2$ (as it turns out, a real effect due to galactic evolution).  

Further, in 1968 Zel'dovich \cite{ZelLambda} pointed out that the energy of the quantum vacuum was mathematically equivalent to a cosmological constant and that it should be enormous, more than 50 orders-of-magnitude greater than the critical density.\footnote{Pauli and others had realized this even earlier, but did not formalize their concerns -- cosmology wasn't mainstream science yet \cite{EPJH}.}  In 1989, Weinberg reviewed the problems of the cosmological constant and suggested -- gulp -- an anthropic solution \cite{WeinbergLambda}.  I believe that
many had the feeling that ignoring $\Lambda$ and making  progress elsewhere was the best strategy.

The COBE detection of CMB anisotropy in 1992 \cite{COBEDMR} added urgency.  It provided a second normalization of the power spectrum of density inhomogeneities and a consistency check, since 
 the shape of the spectrum was fixed to be close to scale-invariant,\footnote{While the inflation-predicted spectrum of density perturbations is scale-invariant in the {\it the gravitational potential,} the transition from the early radiation-dominated era to the matter-dominated era imprints a feature on the spectrum of density perturbations.  That feature, see Fig.~\ref{fig:LSS}, depends upon when that transition took place and scales with $\Omega_Mh$.} In order to fit both, $\Omega_M h \sim 0.3$ was required.  Beyond the two obvious fixes, small Hubble constant ($h\sim 0.3$) or a cosmological constant with $\Omega_\Lambda \sim 0.7$, there were other ways to fix the problem, including adding a bit of hot dark matter or extra radiation \cite{flavorsofCDM}.

The final nail in the coffin of $\Omega_M = 1$ was the cluster inventory determination of the total matter density \cite{ClusterFairSample}.   Assuming that clusters provide a fair sample of matter in the Universe,\footnote{In clusters, most of the baryons are in the hot, x-ray emitting gas, whose mass is determined by the x-ray emission; the total mass comes from virial theorem or weak lensing estimates.} the BBN-determined baryon density and the ratio of baryons-to-total mass measured in clusters determines $\Omega_M$. The result, $\Omega_M (h/0.7)^{1/2} \sim 0.3$, was convincingly less than one, unless the Hubble constant is very small.

In March 1995, Krauss and I boldly proposed that ``the cosmological constant is back," showing that it solved all the problems of inflation + CDM for $\Omega_\Lambda \sim 0.6$ and $h\sim 0.6$ \cite{KraussTurner}.  We were surprised at how little outcry there was from the particle physics community about invoking $\Lambda$.  There was one piece of evidence against our proposal, a 95\% confidence upper limit of $\Omega_\Lambda < 0.66$, based upon the statistics of gravitational lenses \cite{Kochanek}.

At the 1996 {\it Critical Dialogues in Cosmology} meeting, at Princeton, I was asked to make the case for $\Lambda$CDM in a session devoted to the different versions of inflation + CDM.
I noted that the definitive test of $\Lambda$CDM was its prediction of $q_0 \sim -0.5$ rather than $q_0 \sim 0.5$ for the other versions of CDM + Inflation.  In a strange twist of fate, at the same meeting, Perlmutter presented results from the Supernova Cosmology Project's (SCP) first 7 high-redshift ($z = 0.35-0.46$) supernovae:  $\Omega_\Lambda < 0.51$ at 95\% confidence \cite{Perlmutter1997}.

\subsubsection{It's $\Lambda$CDM}
A lot happened in the next two years.  Both teams amassed more type Ia supernovae (42 for the SCP and 44 for the High-z team) and, in early 1998,  announced their results:  The Universe is accelerating, consistent with $\Omega_\Lambda \sim 0.7$ and $\Omega_M \sim 0.3$.    

Remarkably, there was rapid acceptance of their extraordinary claim.  I believe there were two reasons for that:  First, the two very competitive and independent teams came to the same conclusion, each with a very thorough analysis.  And second, $\Lambda$ was the final piece of the puzzle, the one that made everything work.

The acceptance was swift as this story illustrates.  My mentor David Schramm was scheduled to debate Jim Peebles in April 1998 on the question of whether the Universe was flat or not; he had flat.  All fall, Schramm was hoping for last-minute results to make his case viable.  After he died tragically in December 1997, I was asked to fill in.  Peebles was no longer willing to debate flat or not, and we debate, ``Cosmology solved?"  The debate took place in October 1998, and I took ``yes"  \cite{PASPGreatDebate}.

\paragraph{Dark energy}  In 1997, White and I generalized the idea of a smooth component with large negative pressure by introducing the equation-of-state parameter, $w\equiv p/\rho =-1$ for $\Lambda$ \cite{TurnerWhite}, and in August 1998 I introduced the term ``dark energy" to describe the smooth component  that was causing the Universe to accelerate \cite{TurnerDE}.  I did so mindful of the fact that while $\Lambda$ was the simplest example of dark energy and that the data were (and continue to be) consistent with it, $w = -1.026\pm 0.041$ \cite{Scolnicetal2018}, there was no compelling reason to believe that $\Lambda$ was the actual cause.  I remain convinced that cosmic acceleration is the most profound mystery in all of science and the question should remain open to challenge both theorists and observers.


Carl Sagan wisely advised that extraordinary claims require extraordinary evidence.  That the Universe is accelerating is certainly an extraordinary claim!  The extraordinary evidence was not in place in 1998; however, it is today. I mark the turning point as April 2000, when Lange's Boomerang team provided their evidence for a flat Universe from the CMB \cite{Boomerang2000}.  Two-thirds of the critical density was missing, and dark energy in the form of $\Lambda$ was the perfect fit.  Additional strong evidence from a very different direction.

\section{TRIUMPHS AND CHALLENGES}
I began using the term ``precision cosmology" in  the 1990s to announce with some fanfare that in cosmology, the errors are no longer in the exponents and the field is no longer data starved!  While the title of this review suggests precision cosmology is a destination; in fact, it is a milestone of a field challenged by the vastness of its undertaking.  
What enabled precision cosmology?  It is simple: powerful tools, bold ideas and great people.  And a bit of luck.  In what follows, I will discuss its accomplishments and a brief look forward.   

\subsection{Triumphs of $\Lambda$CDM}
\begin{figure}[h]
\begin{subfigure}{0.65\textwidth}
\includegraphics[width=0.95\linewidth, height = 5cm]{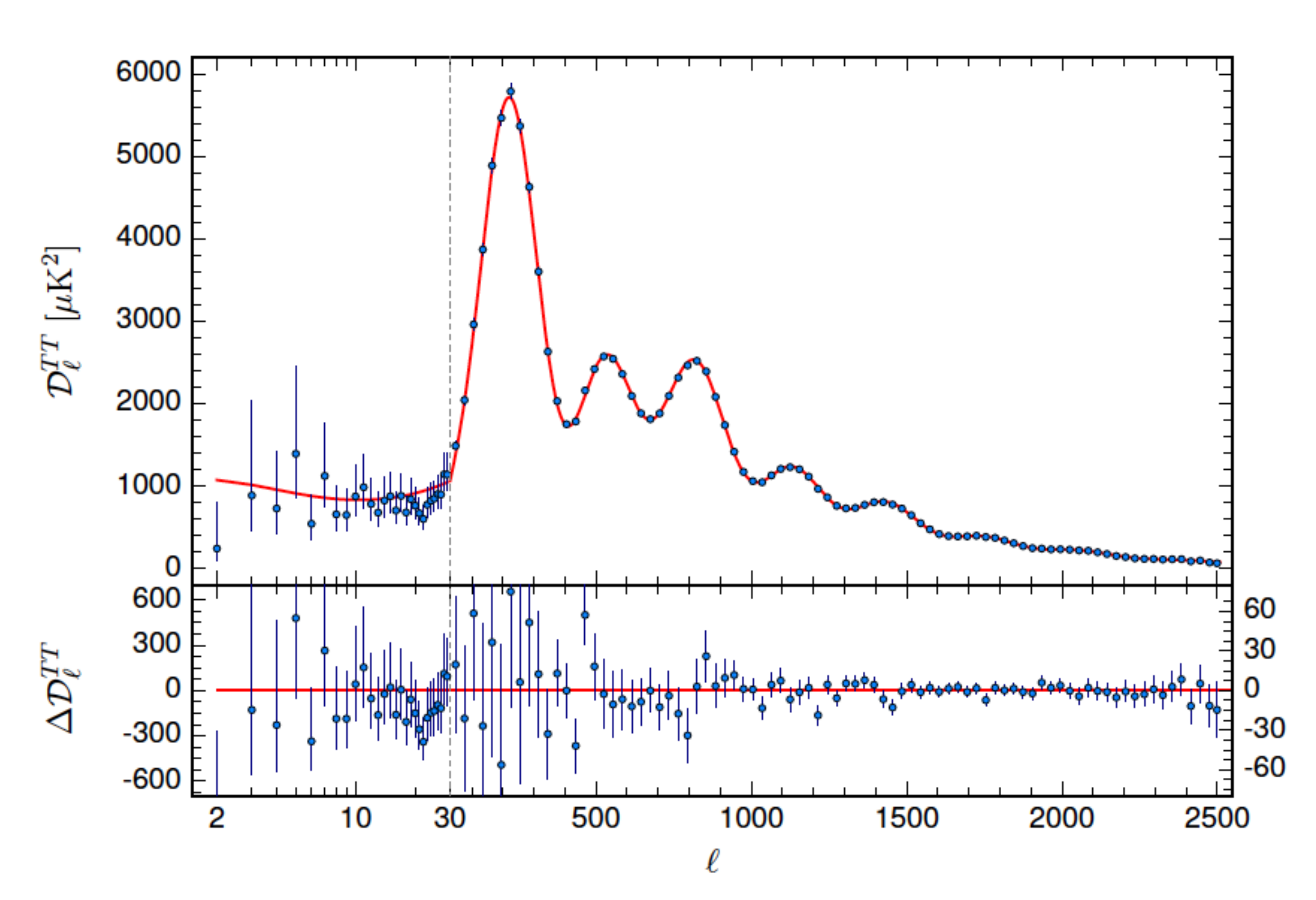}
\end{subfigure}
\begin{subfigure}{0.35\textwidth}
\includegraphics[width=0.95\linewidth, height = 3.5cm]{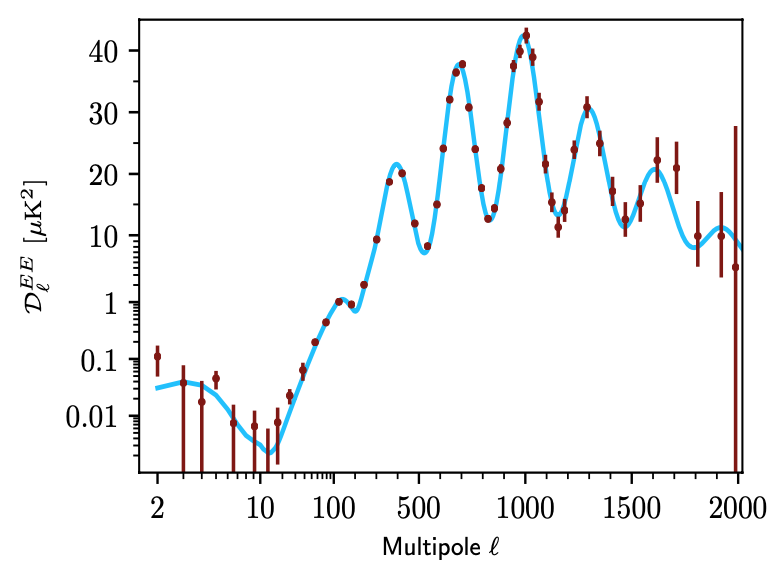}
\end{subfigure}
\caption{Just six numbers describe the Universe. The Planck TT angular power spectrum and residuals from the best fit six-parameter model (left) and EE power spectrum (right).}
\label{fig:PlanckAPS}
\end{figure}

\begin{figure}[h]
\begin{subfigure}{0.55\textwidth}
\includegraphics[width=0.95\linewidth, height = 4.8cm]{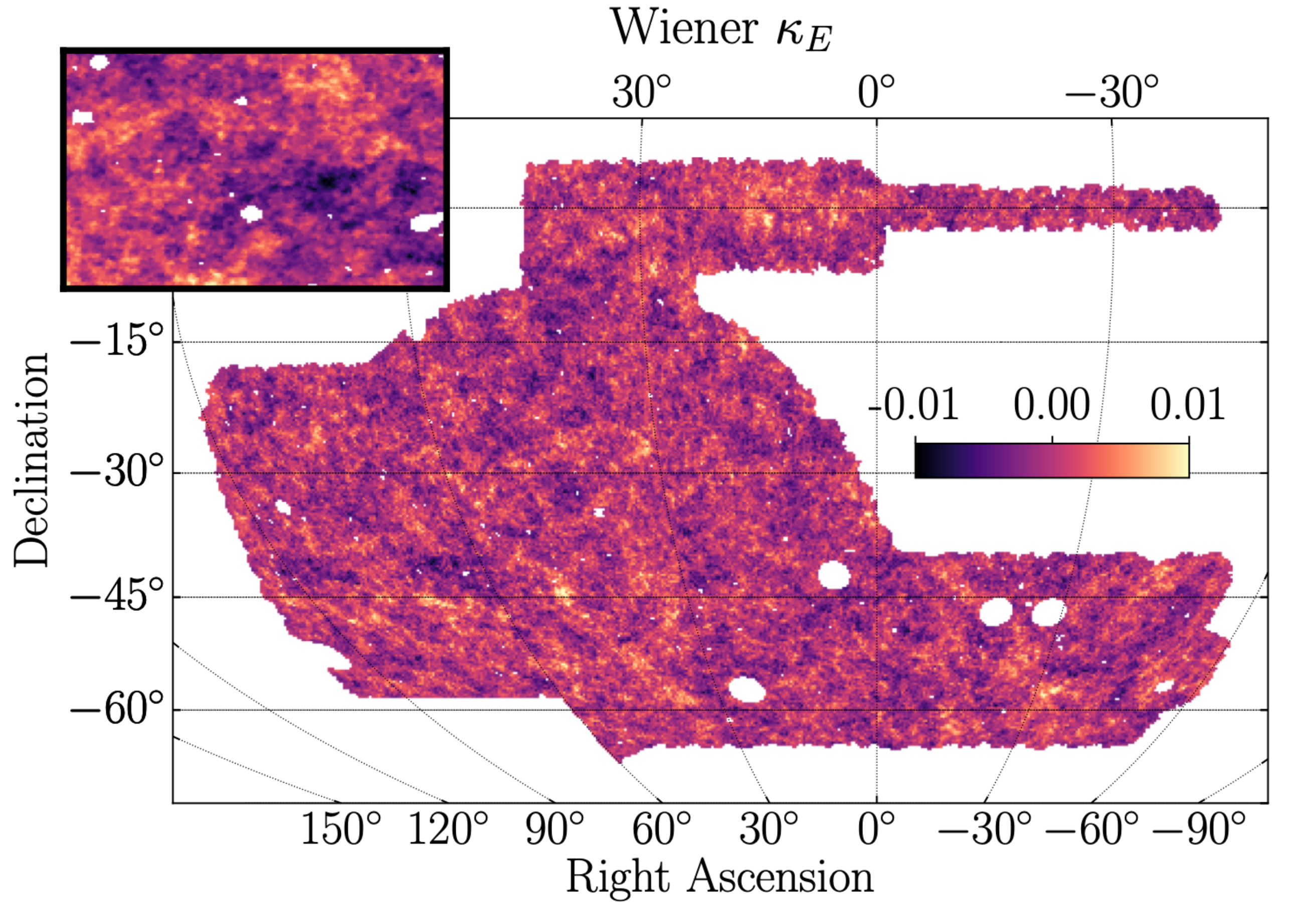}
\end{subfigure}
\begin{subfigure}{0.45\textwidth}
\includegraphics[width=0.95\linewidth, height = 3.8cm]{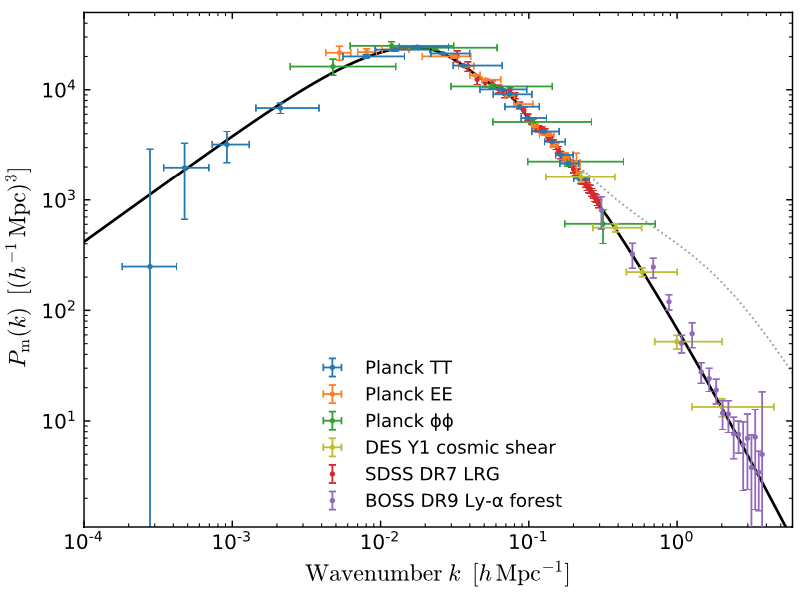}
\end{subfigure}
\caption{Large-scale structure measurements.  left: DES weak-lensing map of 5000 sq. deg. of the sky \cite{DES_WL}.  right:  Planck compilation of the power spectrum of inhomogeneities; the curve is $\Lambda$CDM model \cite{PlanckLegacy}.}
\label{fig:LSS}
\end{figure}

The emergence of $\Lambda$CDM as the second cosmological paradigm marks the first triumph of precision cosmology.  This new paradigm illustrates how far cosmology has come from its early days.  First, in the depth of understanding that it provides about the origin and evolution of the Universe; next, in the impressive body of evidence -- both precision measurements and crosschecks -- upon which it rests; and finally, in the profound set of questions that it poses.
Here is a brief summary of our current understanding:  
\begin{extract}
\vskip -10pt
{\it Inflation:} A very early ($\ll 10^{-5}\,$sec) period of accelerated expansion is driven by the potential energy of a weakly-coupled scalar field and results in a small, smooth patch of the Universe growing exponentially in size to easily encompass all that we see today; quantum fluctuations in the scalar field grow in size and become an almost scale-invariant spectrum of nearly-Gaussian density perturbations that seed all the structure seen today; quantum fluctuations in the metric of spacetime grow in size and become an almost scale-invariant spectrum of gravitational waves; and inflation ends when the scalar-field potential energy is converted into particles that thermalize and initiate the hot ``quark soup" phase.
\vskip 4pt
{\it Quark soup:}  During this phase of mostly thermal-equilibrium, which lasts from the end of inflation ($t\ll 10^{-5}\,$sec) until quarks form into hadrons ($t\sim 10^{-5}\,$sec), two  non-equilibrium events occur: baryogenesis leads to a small excess of baryons over antibaryons, and dark matter particles are created.  Other non-equilibrium processes may have occurred leaving relics yet to be discovered.
\vskip 4pt
{\it BBN:} Nuclear reactions take place from $10^{-2}\,$sec to $200\,$sec, leading to the synthesis of a significant amount of $^4$He, and trace amounts of D, $^3$He, and $^7$Li, which are still with us today as the oldest relics of the early Universe -- at least currently.
\vskip 4pt
{\it Gravity builds cosmic structure:} As the Universe became matter-dominated ($t\sim 64,000\,$yrs) baryons fell into the gravitational potential wells of the CDM particles and owing to the resisting pressure of photons underwent acoustic oscillations until the formation of neutral atoms ($t \simeq 380,000\,$yrs); thereafter, photons streamed freely becoming the CMB and structure formed in a hierarchical manner, from galaxies to clusters of the galaxies to superclusters.  The CMB provides a snapshot of the Universe around last scattering, with a wealth of information encoded in the acoustic peaks and other features.
\vskip 4pt
{\it Cosmic acceleration:} About 5\,Gyr ago, the repulsive of gravity of dark energy -- consistent with a cosmological constant -- overtook the attractive gravity of dark matter, leading to the current epoch of accelerated expansion and inhibiting the formation of larger structures.  

\end{extract}
\vskip -7pt
The body of evidence that supports $\Lambda$CDM includes CMB anisotropy and polarization on angular scales from $90^\circ$ to arcminutes; large-scale structure as quantified by galaxy redshift surveys, weak lensing and baryon-acoustic oscillations (BAO); the expansion history as probed by type 1a supernova and BAO; the light-element abundances; cluster abundances; and observations of galaxy properties and evolution.  Two figures cannot summarize the impressive the vast amount of data; however, see Figs.~\ref{fig:PlanckAPS} and \ref{fig:LSS}.


\subsubsection{Planck speaks} Consider just a small subset of cosmological parameters from the final Planck results, almost all measured with sub-percent precision \cite{PlanckLegacy}: 
\begin{enumerate}
\item{} $\Omega_0 = 0.999 \pm 0.002$
\item{} $n_S = 0.9649 \pm 0.0042$
\item{} $\Omega_B h^2 = 0.02237 \pm 0.00015$
\item{} $\Omega_M h^2 = 0.143 \pm 0.0011$
\item{} $t_0 = 13.80 \pm 0.023\, {\rm Gyr}$
\item{} $H_0 = 67.4 \pm 0.54\,{\rm km/s/Mpc} $
\item{} $\sigma_8 = 0.8111 \pm 0.0060$
\end{enumerate}

The first two numbers, together with the Planck upper limits to any non-Gaussianity \cite{PlanckNG}, provide strong evidence for the basic inflationary predictions of a flat Universe, with almost scale-invariant, Gaussian curvature perturbations.  The power-law index of density inhomogeneity $n_S =1$ corresponds to scale-invariance; the expectation of a small deviation \cite{HT} shows up at almost 10$\sigma$.

Next comes the almost 100$\sigma$ difference between the baryon density and the matter density.  It anchors the case for non-baryonic dark matter, unless something is wrong with the whole framework.  Planck also limits the amount of HDM, $\Omega_\nu h^2 < 0.0013$, adding to the large-scale structure measurements that scream out for cold dark matter. 

The last three cosmological parameters in my list are derived from CMB anisotropy measurements and the assumption of $\Lambda$CDM.
Having an age determination with an error estimate of 23 million years is a far cry from the 1990s when age estimates ranged from 10\,Gyr to 20\,Gyr.  Further, the CMB measurements of $H_0$ and $\sigma_8$ (the level of inhomogeneity on the cluster scale of $8h^{-1}\,$Mpc), allow for end-to-end consistency tests of $\Lambda$CDM, by comparing with direct measurements of  both quantities today; more in~\S4.2.2.


\subsubsection{Deuterium}
\begin{figure}[h]
\begin{subfigure}{0.60\textwidth}
\includegraphics[width=0.95\linewidth, height = 4cm]{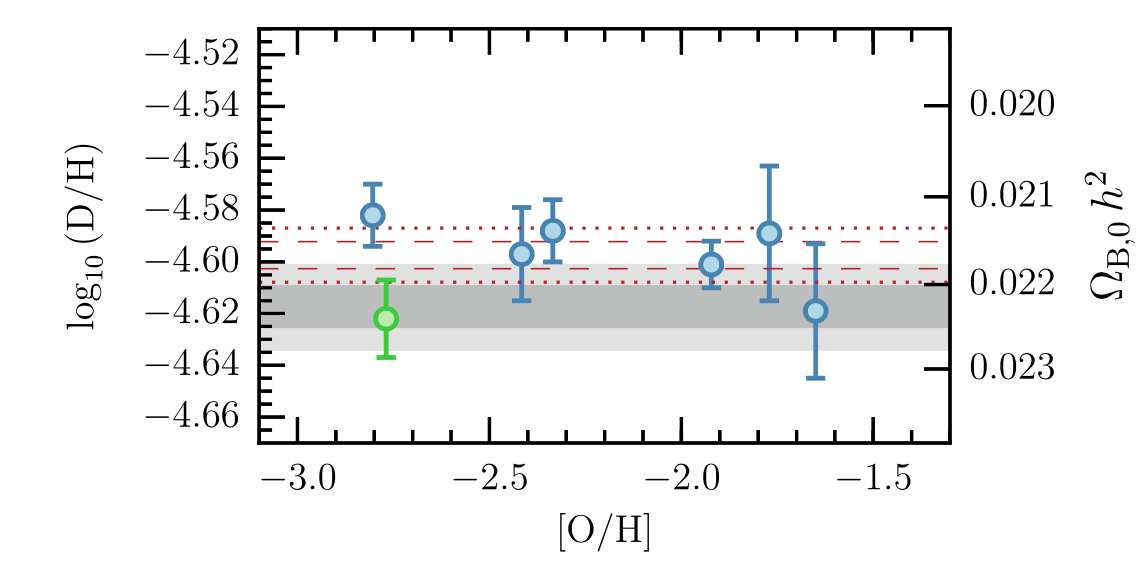}
\end{subfigure}
\begin{subfigure}{0.38\textwidth}
\includegraphics[width=0.95\linewidth, height = 5.8cm]{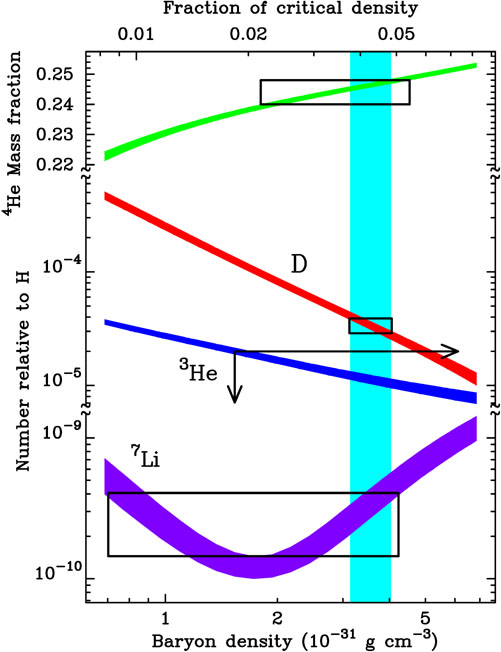}
\end{subfigure}
\caption{Big bang nucleosynthesis.  left:  D/H determinations.  right:  The vertical band is the deuterium-determined baryon density and the other bands are the one-sigma predictions.  The heights of the black boxes indicate the measured abundances with error estimates.}
\label{fig:BBN}
\end{figure}

Big-bang nucleosynthesis played a crucial role in establishing the hot big bang model with its successful explanation of the large primordial abundance of $^4$He; in the precision era, deuterium takes centerstage.  Deuterium it is the ``baryometer" \cite{DNSMST} because its abundance depends strongly upon the baryon density; see Fig. \ref{fig:BBN}. 

In 1996 the primordial abundance of deuterium was determined by measuring the absorption of QSO light by intervening gas clouds of pristine material \cite{Tytler}.  Clouds with just the right properties are few and far between, and it took the high-resolution spectrograph on the 10-meter Keck telescopes to identify the deuterium feature.  With a dozen detections, the precision is approaching 1\%, D/H $=2.527 \pm 0.03 \times 10^{-5}$ \cite{Cooketal}. 

This abundance implies a baryon density, $\Omega_B h^2 = 0.02166 \pm 0.00015 \pm 0.00011$,\footnote{The second uncertainty is the theoretical uncertainty arising from input nuclear data, primarily the cross section for d$(p,\gamma )^3$He; see \cite{YOF}.}  
with precision better than 1\% and consistency with the Planck determination.  The predicted mass fraction of $^4$He that follows from this is $Y_P = 0.2469 \pm 0.0002$, which is consistent with astrophysical measurements, $Y_P = 0.245 \pm 0.0034$.  Further, both are consistent with the less-precise Planck-determined $^4$He abundance, $Y_P = 0.242 \pm 0.024$, based upon the structure of the acoustic peaks.

The baryon density is a poster child for precision cosmology:  two, percent-level determinations of a fundamental quantity that agree. One derives from gravity-driven acoustic oscillations when the Universe was 380,000 yrs old, while the other involves nuclear reactions that took place when the Universe was seconds old.  Wow!  

\subsubsection{Loose ends} With precision comes controversy.  The biggest one involves $^7$Li, whose predicted abundance is a factor of two larger than the abundance seen in the lowest-metallicity stars, a blemish on an otherwise stellar record for BBN.  There is a growing consensus that the putative primordial abundance, which traces back to the 1980s, was a mirage and actually reflected stellar depletion of the primordial value \cite{Lithium}.  Stay tuned.

From the beginning, cold dark matter has been ``under attack" -- as any attractive and expansive theory should be  \cite{CDMproblems}.  Mostly, the problems involve small scales (e.g., predicted shapes of the central rotation curves of dwarf galaxies, or the underabundance of dwarf galaxies as compared to observations), where the influence of baryons {\it can be} significant and {\it can possibly} explain the discrepancy. I would be more concerned if the problems involved larger scales, where baryonic hydrodynamics cannot be important.  
Having witnessed at least ``nine lives" of CDM, I believe that it contains most of the truth about structure formation,  and if a replacement is needed, it will  look very similar to CDM.

\subsubsection{Cosmic limits}
Cosmology is not laboratory science.
We are constrained to observe rather than set up experiments, and we cannot re-create the big bang or try alternative cosmologies or scenarios.
We are also creatures of our own limited time and space horizons.  The ``standard of proof" in cosmology must also be different.   I -- and I think most other cosmologists -- are convinced that big bang nucleosynthesis {\it really happened},  because we use physics well-established in the laboratory to make detailed predictions that agree with observations.  What will we require to be able to say the same for dark matter or inflation?

\subsubsection{The heavenly lab}
The other big triumph of the coming together of  ``the very big" and  ``the very small" is the ability to make ``precision measurements" in the heavenly laboratory.  It has opened the door to a new way of making discoveries and testing theories.

Historically, 
the first such discovery is that of the element Helium, through its absorption lines in the sun more than 100 years ago.  The list now includes neutrino mass (solar neutrinos and atmospheric neutrinos),  3 generations of quarks and leptons (BBN $^4$He production), a new stable particle (dark matter), non-particulate stress-energy with repulsive gravity (dark energy), and the existence of gravitational waves that propagate with the speed of light (LIGO/Virgo detection of gravitational waves).  

Today, setting limits to hypothetical particles, objects, events, and on and on has become a well accepted and trusted way to extend the reach of terrestrial laboratories when testing new ideas in fundamental physics.

\subsection{Toward a third cosmological paradigm}
Two big thinkers -- Niels Bohr and Yogi Berra -- agree that prediction, especially about the future, is hard.
Add to this, cosmology is boom or bust science, with long intervals between  big discoveries:  40 years from Hubble to Penzias and Wilson; 27 years to the COBE discovery of anisotropy; 6 years to the discovery of dark energy; and since, 24 years of quiet, steady progress. 

Nevertheless, I can't resist sharing a few thoughts about the cosmological paradigm\footnote{In \cite{thirdparadigm} I introduced ``the third cosmological paradigm" and opined upon it at length.} that may follow $\Lambda$CDM.  The only clarity I have is in the organization of those thoughts:  What we might expect from linear progress, from a Kuhnian-type shift, and from thinking about lofty issues.  

\subsubsection{Known mysteries}
$\Lambda$CDM leaves cosmology with three big questions to answer, and by attempting to do so, make linear progress.  They are:  What is the dark matter particle?  What is the nature of dark energy? and What is the physics underlying inflation?  And there is an impressive program in place to address these questions, from more than ten large-scale searches for dark matter particles to precision probes of cosmic acceleration on the ground and in space to CMB experiments to detect the B-mode polarization signature of inflation.  Surely,  clues or even a big payoff will come from these efforts.

The is another big challenge: a standard model of baryogenesis with consequences that can be tested.  That baryogenesis involves Standard Model physics and the electroweak phase-transition has all but been ruled out.  Viable models now involve physics at scales well beyond those accessible in terrestrial laboratories.  I find the idea that the baryon asymmetry arose as a  lepton asymmetry amongst the neutrinos and was transmuted into a baryon asymmetry by Standard Model $B+L$ violating interactions, but that idea is equally difficult to test \cite{leptogenesis}.


As good as our questions are, they may be the wrong ones.
Dark energy could be a mirage with the real explanation for cosmic acceleration being a replacement for General Relativity.  The dark-matter hunt has focussed in on a handful of well-motivated particles including the neutralino and axion, but the story could be much more complicated, e.g., with dark matter being a portal to a whole dark sector \cite{DarkSector}.  

Inflation in its current form is {\it at best} a first approximation to a more fundamental theory.  Its loose ends -- prediction of a multiverse, uneasy relationship to initial conditions, and the fact that its exponential expansion is the ``microscope" that reveals scales that began much smaller than the Planck scale \cite{AAMicroscope} -- may help illuminate the path forward.

Or, inflation could just be way off the mark.  Early in my career, I saw the quark model evolve from $SU(3)$ flavor to the $SU(3)$ color gauge theory of QCD; the threes could hardly be more different.

We could be on the wrong track, with dark matter, dark energy and inflation being the modern equivalents of Ptolemy's epicycles.  $\Lambda$CDM is complicated with photons, baryons, neutrinos, cold dark matter and dark energy, with unexplained ratios between the matter components \cite{CarrTurner}:  CDM-to-baryons-to-neutrinos, $5.4:1: (0.024 - 0.06)$.  Too much works for us to be completely off track, but there could be an unexpected Kuhnian shift ahead.


\subsubsection{Kuhnian shift}
The place to look for signs of such a shift are  the ``tensions," 2 to 3$\sigma$ discrepancies of independent measurements of the same quantity within $\Lambda$CDM, 
that could grow in significance, and result in a  Kuhnian paradigm shift.

The most urgent one today is the Hubble tension:  direct measurements of the current expansion rate yield $H_0 = 74\pm 1\,$km/sec/Mpc and ``indirect'' measurements of the expansion rate using CMB anisotropy and the {\it assumption} of $\Lambda$CDM to extrapolate the early time expansion rate to the present yield $H_0 = 67.5\pm 0.4 \,$km/sec/Mpc \cite{hubbletension1,hubbletension2}; see Fig.~\ref{fig:H0}.

\begin{figure}[h]
\begin{subfigure}{0.59\textwidth}
\includegraphics[width=0.95\linewidth, height = 4.7cm]{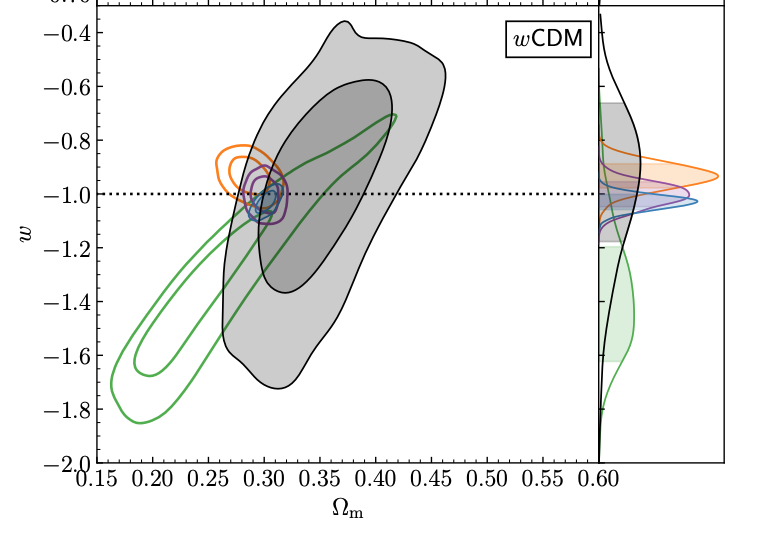}
\end{subfigure}
\begin{subfigure}{0.41\textwidth}
\includegraphics[width=0.95\linewidth, height = 4.9cm]{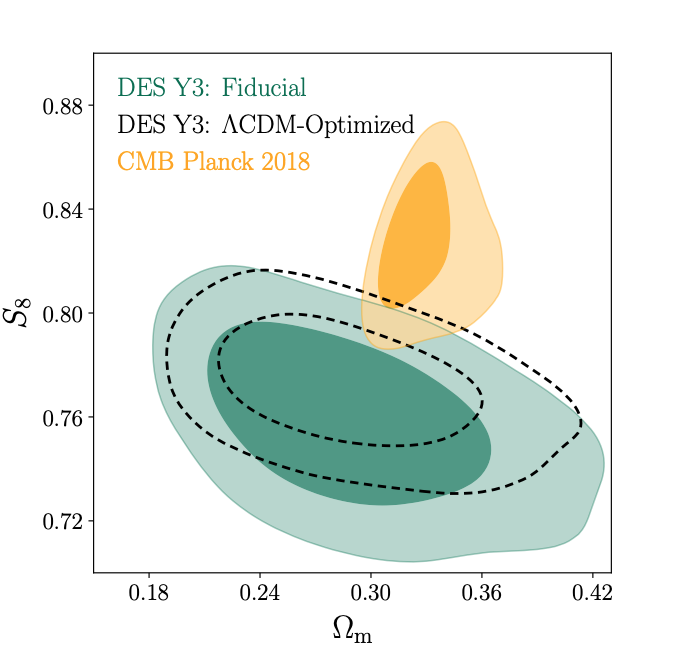}
\end{subfigure}
\caption{Joint analyses of the cosmological parameters $\Omega_M$, $S_8 \equiv \sigma_8 (\Omega_M /0.3)^{0.5}$ 
and dark energy equation-of-state $w$ \cite{sigma8tension}. 
Note the consistency of the Planck and DES precision measurements of $w$ (left), and the slight $S_8$ ``tension" (right).}
\label{fig:DES}
\end{figure}

The resolution could be a systematic error in one (or both) determinations of $H_0$, or, something important missing from $\Lambda$CDM.  At the moment, there is no compelling modification of $\Lambda$CDM that easily accommodates the difference \cite{hubbletensionsolutions}.  Another lingering tension is the discrepancy between direct measurements of $\sigma_8$ and $\Omega_M$ with measurements that come from CMB anisotropy, as shown in Fig.~\ref{fig:DES}. 

\subsubsection{Two lofty issues} 
A tiny discrepancy could illuminate the path forward.  On the other hand, simply asking the correct big question may open the door. Because each brings so much with it, I see two really big issues in cosmology, the big bang event and the multiverse.

According to General Relativity, the big bang is the singular beginning of it all, matter, energy, space and time. It also begs us to ask about initial conditions, are they relevant or not; inflation seeks to make them irrelevant in practice, but in principle.  And while the big bang is a singularity within GR, it is likely not in its successor theory.  This opens the scientific discussion of what happened before the big bang, or the origin of space and time.  It could be that GR has the right answer, but without the supporting mathematical details.  Or, what we think of as the big-bang is actually the portal to an earlier phase of a cyclic universe, a decades-old idea that continues to attract attention.

I am not a fan of the multiverse.  However, the issues it raises are too important to be ignored.  The simplest question we can ask about the Universe -- how  big is it? -- cannot be answered because of the limitations of our past light cone \cite{GFREllis}. The disconnected pieces of a multiverse raise the stakes:  we cannot know how typical the part that we exist in is.   Steinhardt has argued that the multiverse, which is born of inflation, calls into question whether or not inflation makes firm predictions that can be tested, and he concludes that an alternative to it is needed \cite{PJS1,PJS2}.  

In addition to  calling attention to a shortcoming of inflation,  the multiverse is also at the nexus of string theory and cosmology.  And it provides a framework for discussing an anthropic/ensemble view of reality, as well as an explanation for the smallness of the cosmological constant and cosmic acceleration \cite{JoeP}.  While early on string theory and cosmology had the makings of a marriage made in ``the heaven," the search for connections between the two has raised more questions and doubts than answers and home runs \cite{SWAMPLAND}.

\subsubsection{Some firm predictions}
As I have demonstrated above, theorists, like economists, are famous for ``on the one hand, on the other hand, ..." style predictions.  With that low bar, I end with some firm predictions.
\begin{enumerate}
\item{}  Cosmology and particle physics will continue to have profound connections.
\item{}  The third cosmological paradigm will look a lot like $\Lambda$CDM; and if we are lucky, with a few surprises.
\item{}  There is a good chance that we will discover a WIMP-like dark-matter particle or dark-matter axions.  If not, all bets are off -- I don't see another compelling candidate.
\item{}  I give 2-to-1 odds that the Hubble tension is resolved without adding something new to $\Lambda$CDM; take heart, 33\% for something new is a really bullish prediction.
\item{}  I am dreaming of the discovery of a gravitational-wave signature in the CMB polarization.  What a window to inflation and the earliest moments it would provide.  I am enough of a realist to know that I may well have my heart broken.
\item{}  I hope for clues about dark energy, but believe this is a truly profound problem and $\Lambda$ may well be a placeholder for many decades.
\item{}  I am not bullish on progress toward a testable model of baryogenesis or an ``up or out" for the multiverse; both are likely to be with us for quite a while.

\end{enumerate}

\subsection{Personal reflections}

It is  impossible to filter out the effects of one's own personal lens in viewing the world.  And in science, it is easy to think that one's career coincided with a special time.  I do feel that the last 50 or so years  has seen especially dramatic progress in cosmology.   The audacious idea that there might be a connection between the very big and the very small has led to revolutionary advances in our understanding of the Universe, and with it, the field has grown from around 30 astronomers to one that comprises 1000s of physicists and astronomers and is front and center on the scientific agendas of both fields.  While I used to take joy in understanding everything that was going on and thinking that I could contribute on virtually every topic, that is no longer possible.  But so much else is.

In 1970, I suspect few imagined that the big questions would evolve so quickly from determining $H_0$ and $q_0$ to  identifying  the dark-matter particle, understanding  dark energy, and unraveling how quantum fluctuations become the seeds for all cosmic structure, or that cosmology would attract the interests of particle physicists. It has been my great  fortune to be part of this grand adventure and of the quirky group that the tribe of cosmologists is. It would be hard to ask for more, but I hope that even more stunning advances and surprises lie ahead for the next generation of cosmologists.

My advice for the next generation of cosmologists: find a scientific question you are passionate about and a style of science you are comfortable with and go at it.  There is plenty to do, from thinking about a really big question and trying to imagine how to make progress to getting involved in a large collaboration trying to detect dark matter particles or probing dark energy.  And be prepared to be flexible; there will be surprises, discoveries about the Universe and discoveries about what captures your passion.


Let me finish by saying that cosmology is a science that does not suit everyone -- having to observe rather than set up experiments, the daunting challenge of the vastness of the Universe and the boom or bust progress.  And more than a little arrogance is required for creatures that evolved from quantum fluctuations and quark soup, that only exist for a short time and are stuck on a small backwater outpost to think that they might be able to understand the whole shebang.  
Steven Weinberg said it more eloquently in the epilogue of the 1993 edition of {\it The First Three Minutes}, ``the effort to understand the universe is one of the very few things which lifts human life a little above the level of farce and gives it some of the grace of tragedy."

\vskip-15pt
\section*{DISCLOSURE STATEMENT}
The author is not aware of any affiliations, memberships, funding, or financial holdings that
might be perceived as affecting the objectivity of this review. 

\section*{ACKNOWLEDGMENTS}
I am most grateful for the comments and illuminating conversations I have had about this review with ``full list to come."
Over the years my research has been generously supported by the DOE Office of High Energy Physics, Fermilab, the NSF Physics Frontier Center program, NASA's Astrophysics Theory program and Office for Innovative research, and the Kavli Foundation.  I also had the good fortune to have the support of both the University of Chicago and Fermilab, and most importantly my mentor, David N. Schramm, who brought me to Chicago and introduced me to cosmology.

%





\end{document}